\documentclass[fleqn,usenatbib]{mnras}

\usepackage{newtxtext,newtxmath}

\usepackage[T1]{fontenc}
\usepackage{ae,aecompl}

\usepackage{graphicx}	
\usepackage{amsmath}	
\usepackage{amssymb}	

\newcommand{\nuCHIME}{\nu_{600{\rm MHz}}}
\defcitealias{Metzger+19}{MMS19}

\newcommand{\be}{\begin{equation}}
\newcommand{\ee}{\end{equation}}

\title[Applying the decelerating blast wave model to FRB populations]{Constraints on the Engines of Fast Radio Bursts}

\author[Margalit, Metzger, \& Sironi]{
Ben Margalit$^{1}$\thanks{NASA Einstein Fellow}\thanks{E-mail: benmargalit@berkeley.edu}, 
Brian D.~Metzger$^{2,3}$, 
and Lorenzo Sironi$^{2}$
\\
$^{1}$Astronomy Department and Theoretical Astrophysics Center, University of California, Berkeley, Berkeley, CA 94720, USA\\
$^{2}$Columbia Astrophysics Laboratory, Columbia University, New York, NY 10027, USA\\
$^{3}$Center for Computational Astrophysics, Flatiron Institute, New York, NY 10010, USA\\
}

\date{Accepted XXX. Received YYY; in original form ZZZ}

\pubyear{2019}

\begin{document}
\label{firstpage}
\pagerange{\pageref{firstpage}--\pageref{lastpage}}
\maketitle

\begin{abstract}
We model the sample of fast radio bursts (FRB), including the newly discovered CHIME repeaters, using the decelerating synchrotron maser blast wave model of \citet*{Metzger+19}, which built on earlier work by \citet{Lyubarsky14}, \citet{Beloborodov17}.  This model postulates that FRBs are precursor radiation from ultra-relativistic magnetized shocks generated as flare ejecta from a central engine collides with an effectively stationary external medium.  Downward drifting of the burst frequency structure naturally arises from the deceleration of the blast-wave coupled with the dependence of the maser spectral energy distribution, and induced Compton scattering depth, on the upstream medium.  The data are consistent with FRBs being produced by flares of energy $E_{\rm flare} \sim 10^{43}-10^{46}(f_{\xi}/10^{-3})^{-4/5}$ erg, where $f_{\xi}$ is the maser efficiency, and minimum bulk Lorentz factors $\Gamma \approx 10^2-10^3$, which generate the observed FRBs at shock radii $r_{\rm sh} \sim 10^{12}-10^{13}$ cm.   We infer upstream densities $n_{\rm ext}(r_{\rm sh}) \sim 10^{2}-10^{4}$ cm$^{-3}$ and radial profiles $n_{\rm ext} \propto r^{-k}$ showing a range of slopes $k \approx [-2,1]$ (which are seen to evolve between bursts), both broadly consistent with the upstream medium being the inner edge of an ion-loaded shell released by a recent energetic flare.  The burst timescales, energetics, rates, and external medium properties are consistent with repeating FRBs arising from young, hyper-active flaring magnetars, but the methodology presented is generally applicable to any central engine which injects energy impulsively into a dense magnetized medium. Several uncertainties and variations of the model regarding the composition and magnetization of the upstream medium, and the effects of the strong electric field of the FRB wave (strength parameter $a \gg 1$) on the upstream medium and its scattering properties, are discussed.  One-dimensional particle-in-cell simulations of magnetized shocks into a pair plasma are presented which demonstrate that high maser efficiency can be preserved, even in the limit $a \gg 1$ in which the FRB wave accelerates the upstream electrons to ultra-relativistic speeds. 
\end{abstract}




\section{Introduction}

Fast radio bursts (FRB) are luminous pulses of radio emission lasting less than a few milliseconds with large dispersion measures (DM) suggestive of an extra-galactic origin (\citealt{Lorimer+07,Keane+12,Thornton+13,Spitler+14,Ravi+15,Champion+16,Petroff+16,Lawrence+17,Shannon+18}). 
The observed short timescale sub-structure in FRB light curves, down to at least tens of microseconds \citep{Michilli+18}, point to their central engines being stellar-mass compact objects such as pulsars \citep{Cordes&Wasserman16,Munoz+19} or magnetars (e.g.~\citealt{Popov&Postnov13,Lyubarsky14,Kulkarni+15}), especially at particularly active stages in their lives \citep{Metzger+17,Beloborodov17}.  

The first confirmed cosmological burster was the repeating source FRB~121102 \citep{Spitler+14,Spitler+16,Scholz+16,Law+17} and its localization \citep{Chatterjee+17} to a dwarf star-forming galaxy \citep{Tendulkar+17}.  FRB~121102 is spatially coincident with a 
compact unresolved ($< 0.7$ pc) luminous persistent synchrotron radio source \citep{Chatterjee+17,Marcote+17}.  The bursts also possess an enormous rotation measure, RM $\sim 10^{5}$ rad m$^{-2}$ \citep{Michilli+18}.  The persistent emission and high-RM likely originate from the same medium, showing that the FRB source is embedded in a dense magnetized plasma (e.g.~\citealt{Michilli+18,Vedantham&Ravi19,Gruzinov&Levin19}).  While this environment could be unrelated to the bursting source itself (for instance, a flaring magnetar that just happens to reside close to an AGN; \citealt{Eatough+13}), it could instead be a compact transient nebula of hot plasma and magnetic fields powered by the FRB central engine \citep{Murase+16,Metzger+17,Beloborodov17,Waxman17}.  

The CHIME survey recently reported the discovery of nine new repeating FRBs \citep{CHIME_R2,CHIME_newrepeaters}.   These repeaters span a wide range of DM $\sim 104-1281$ pc cm$^{-3}$, including one burster with a low DM value (below the inferred Galactic maximum along the line of sight) and RM $\simeq -114.6$ rad m$^{-2}$.  All but one of the CHIME repeaters also exhibit a downward drift in the frequency structure of sub-bursts over the course of the burst duration, intriguing behavior remarkably similar to that first observed in FRB~121102 \citep{Gajjar+18,Hessels+19}. 

Despite extensive monitoring, most FRBs have been observed to burst only once (e.g.~\citealt{Petroff+15}).  However, because of selection effects and uncertainty about how representative the properties of FRB 121102 are of repeaters in general, it remains an open question whether the apparently non-repeating class form a physically distinct population  (e.g.~\citealt{Caleb+18,James19}).  The burst arrival times from FRB~121102 are non-Poissonian and clustered (e.g.~\citealt{Spitler+14,Spitler+16,Opperman+18,Katz18,Li+19}), with extended ``dark'' phases of little or no apparent FRB activity (e.g.~\citealt{Price+18}).  Two additional bursts were recently detected from one of the brightest ASKAP bursts, FRB 171019 \citep{Kumar+19}, showing that at least some in this population can repeat.  \citet{Lu&Piro19} argue that the DM distribution of the ASKAP sample is consistent with a burst luminosity function matching that of repeating FRBs.  Furthermore, the DM distribution of the CHIME repeating FRB sample is statistically indistinguishable from those which have not yet been observed to repeat \citep{CHIME_newrepeaters}.  On the other hand, the widths of FRB~121012's bursts \citep{Scholz+16} and the CHIME repeaters \citep{CHIME_newrepeaters} may be systematically longer than those of the non-repeating, single-component FRBs.

Distinct classes of FRBs would be expected to manifest through the demographics of their host galaxies.  While host associations have not yet been reported for the CHIME FRBs, two of the published ASKAP bursts have been well-localized, FRB 180924 \citep{Bannister+19} and FRB 181112 \citep{Prochaska+19}.  The host galaxies of both bursts appear to be more massive than the dwarf host of FRB~121102.  The Deep Synoptic Array ten-antenna prototype (DSA-10) also recently localized an FRB to a few arcsecond region, consistent with the position of a massive galaxy at $z = 0.66$ \citep{Ravi+19}. Despite growing evidence for a diversity of large-scale environments, it still remains possible that most FRBs occur in regions of active star formation.      

The enormous brightness temperatures $\gtrsim 10^{37}$ K of FRBs demand a coherent emission process (e.g.~\citealt{Katz16,Lyutikov19}).  A number of models explored in the literature postulate FRB emission to take place in the magnetosphere of a neutron star, e.g. through the curvature emission mechanism (e.g.~\citealt{Cordes&Wasserman16,Kumar+17,Katz18b,Lu&Kumar18,Yang&Zhang18,Wadiasingh&Timokhin19,Wadiasingh+19,Lyutikov19}).    

Here we instead continue our focus on a model in which FRB emission is generated on much larger spatial scales from the central engine, by the maser synchrotron process (e.g.~\citealt{Hoshino&Arons91,Long&Peer18}) in an ultra-relativistic shock moving towards the observer into a magnetized medium \citep{Lyubarsky14,Beloborodov17,Plotnikov&Sironi19}.  Larmor rotation of charges entering the shock and gyrating around the ordered magnetic field creates the necessary population inversion in the form of an unstable ring-like particle distribution function, which relaxes by transferring energy into an outwardly propagating coherent electromagnetic wave (e.g.~\citealt{Gallant+92,hoshino_92,Amato&Arons06,Hoshino08,Sironi&Spitkovsky09,Sironi&Spitkovsky10,Iwamoto+17,Iwamoto+18,Plotnikov&Sironi19}).  

The synchrotron maser can naturally account for the high measured linear polarization of some FRBs \citep{Ravi+16,Petroff+17,Caleb+18}, which for FRB 121102 is nearly 100\% \citep{Michilli+18,Gajjar+18} and is also high for at least one of the CHIME repeaters \citep{CHIME_newrepeaters}.\footnote{Some FRBs show no detectable linear polarization.  However, this may be the result of propagation effects in a local magnetized medium, such as Faraday rotation (which cannot be subtracted off without sufficient spectral resolution; \citealt{Michilli+18}) or Faraday conversion into circularly polarized emission \citep{Vedantham&Ravi19,Gruzinov&Levin19}.}   Particle-in-cell (PIC) simulations show that a large-amplitude linearly polarized X-mode wave (the nascent FRB) is created at the shock front by the synchrotron maser instability and propagates into the upstream medium (e.g.~\citealt{Gallant+92}).   

Another favorable aspect of the maser is its potentially high efficiency, $f_{\xi}$, for converting the kinetic energy of the flare into coherent electromagnetic radiation.  One-dimensional PIC simulations of shocks propagating into an electron/positron pair plasma find a maximum efficiency of up to several percent for an upstream magnetization\footnote{Here the magnetization is defined as the ratio of the incoming upstream Poynting flux to the kinetic energy flux.} $\sigma \sim 0.1$, which decreases as $f_{\xi} \propto \sigma^{-2}$ for $\sigma \gg 1$ \citep{Plotnikov&Sironi19}.  The efficiency is lower in the physical case of higher dimensions \citep{Sironi&Spitkovsky09}, but only by a factor of $\lesssim 10$ for $\sigma\lesssim 1$ (\citealt{Iwamoto+17,Iwamoto+18}) and less at high magnetization $\sigma\gtrsim 1$ (\citealt{Plotnikov&Sironi19}; Sironi et al, in prep).  The efficiency is less understood when the upstream medium possesses an electron-proton or electron-positron-proton composition; current PIC simulations in these cases are mostly limited to 1D setups (\citealt{hoshino_92,Amato&Arons06,Hoshino08,stockem_12}; but see \citealt{Iwamoto+19} for 2D simulations) which have not convincingly reached a steady state in the precursor FRB emission.  Another issue of potential importance is the back-reaction of the precursor wave electric field on the upstream plasma, which in the limit of a strong wave is expected to accelerate the upstream electrons to ultra-relativistic speeds before they enter the shock (e.g.~\citealt{lyubarsky_06,Lyubarsky19}).

If the central engine is a flaring magnetar, a relativistic blast wave may be generated by the transient release of energy from the magnetosphere during the initial stage of a flare, which then transforms into an outgoing $\sigma \gg 1$ pulse that collides and shocks the surrounding environment, generating the FRB on much larger radial scales \citep{Lyubarsky14,Beloborodov17}.  The properties of a given burst are then a function of both the engine output (flare duration, energy) and the surrounding medium (density, composition, magnetization).  

Clues to the large-scale environment surrounding the engine are provided by the RM and persistent source flux.  \citet{Margalit&Metzger18} show that a single expanding and continuously-energized magnetized ion-electron nebula embedded within a decades old supernova remnant is consistent with all of the properties of the persistent source of FRB~121102 (size, flux, self-absorption constraints) and the large but decreasing RM (see also \citealt{Margalit+18,Wang&Lai19}).  
The high time-averaged baryon flux into the nebula needed to explain the persistent source and RM of FRB~121102 contrasts with the much ``cleaner'' but short-lived $\lesssim 1$~ms ultra-relativistic ejection events needed to power FRBs themselves.  This suggests a picture, first proposed by \citet{Beloborodov17}, in which the bulk of the ejected mass emerges from the central engine after major flares, and then subsequently serves as the upstream medium into which the next flare collides to produce the FRB.  

Informed by the spectral energy distribution (SED) of the shock precursor predicted by PIC simulations, \citet*[hereafter \citetalias{Metzger+19}]{Metzger+19} calculated the time-dependent synchrotron maser emission
 from a decelerating ultra-relativistic blast wave propagating into a slowly-expanding upstream medium.  Given an average ion injection rate $\sim 10^{20}-10^{21}$ g s$^{-1}$ and intervals between major ion ejection events $\sim$ 1 d similar to the occurrence rate of the most powerful bursts from FRB~121102 and consistent with the statistics of the CHIME repeaters, they demonstrated the production of $\sim$GHz bursts with isotropic radiated energies $\sim 10^{37}-10^{40}$ erg and durations $\sim 0.1-10$ ms for flare energies $\sim 10^{43}-10^{45}$ ergs.  They further showed how the deceleration of the blast wave, and increasing transparency of the upstream medium to induced Compton scattering, naturally generates temporal decay of the observed burst frequency, similar to the observed downward frequency drift of FRB sub-pulses \citep{Gajjar+18,Hessels+19,CHIME_R2,CHIME_newrepeaters}.   

Motivated by recent discoveries by CHIME, here we bring the \citetalias{Metzger+19} model to bear on the newly expanded FRB sample.  Our goals are both to check the assumptions of the decelerating blast-wave scenario and, given success in this regard, to draw inferences about a {\it population} of FRB central engines.  Although a magnetar engine is appealing for many reasons, 
there may be alternative proposed models that
postulate the impulsive injection of energy into a dense and potentially magnetized external environment.  We therefore attempt to keep our modeling independent of the specifics of the engine, returning to this issue in the discussion.

In $\S\ref{sec:model}$ we summarize the key results of \citetalias{Metzger+19} and apply them to the new repeater sample to derive properties of the FRB-producing shocks and the upstream medium.  In $\S\ref{sec:variations}$ we describe uncertainties of the model, which we partially address by presenting the results of PIC simulations of synchrotron maser emission from magnetized shocks in the strong-wave limit $a \gg 1$.  We then apply our results from $\S\ref{sec:model}$ to draw inferences about the nature of the FRB central engines ($\S\ref{sec:engine}$) and the upstream medium ($\S\ref{sec:upstream}$).  \citet{Beloborodov19} recently criticized aspects of the \citetalias{Metzger+19} model, particularly regarding the assumption of the upstream medium being a slowly-expanding ion shell and our treatment of induced Compton scattering in the $a \gg 1$ limit.  \citet{Beloborodov17} and \citet{Beloborodov19} instead advocate that the upstream medium into which the magnetic flares collide is an ultra-relativistic rotationally-powered component of the magnetar wind between flares.  We address these concerns in $\S\ref{sec:variations}$ and provide an explicit comparison between our models in $\S\ref{sec:Beloborodov}$.

\section{FRBs from Decelerating Blast-waves}
\label{sec:model}

\citetalias{Metzger+19} consider the general scenario of maser emission from a decelerating ultra-relativistic blast wave, created by the impulsive injection of flare energy $E_{\rm flare}$ of duration $\delta t$ into a slowly expanding (effectively stationary) external medium with a power-law radial density profile $n_{\rm ext} \propto r^{-k}$ and magnetization $\sigma \lesssim 1$.  Here, we review these predictions and apply them to the recently expanded sample of repeating and (thus far) non-repeating FRB sources.  A secondary goal of this section is to provide an easy-to-use template for applying the model to future expanded FRB samples.

\subsection{Downward Drifting Frequency}

We begin with the downward frequency drifting of the burst sub-pulses seen in FRB~121102 \citep{Hessels+19} and the CHIME repeaters \citep{CHIME_newrepeaters}. For an upstream medium of moderate magnetization $\sigma \lesssim 1$, the SED of the synchrotron maser peaks in the comoving post-shock frame at the frequency $\nu'_{\rm pk} \approx 3\nu_{\rm p}$, where $\nu_{\rm p}$ is the plasma frequency of the upstream medium, with power extending to frequencies $\gg \nu_{\rm p}$ and detailed structures due to overlapping line-like features produced by a large number of resonances \citep{Plotnikov&Sironi19}.  Sub-structure is also imprinted in the SED when the upstream electrons have a finite temperature, which tends to suppress the high frequency harmonics (Babul \& Sironi, in prep).  These spectral features are at least qualitatively consistent with the observed complex, and sometimes narrow-band SEDs, of observed FRBs (e.g.~\citealt{Ravi+16,Law+17,Macquart+19}).  

The peak frequency as seen by an observer ahead of the shock is higher than the rest-frame $\nu'_{\rm pk}$ by a factor $\Gamma \gg 1$, where $\Gamma$ is the bulk Lorentz factor of the shocked gas.  The shock decelerates as it sweeps up mass, such that $\Gamma$ decreases in time.  Combined with the evolution of $\nu'_{\rm pk} \propto n_{\rm ext}^{1/2} \propto r^{-k/2}$, this causes the SED to drift downwards in frequency.  In such a picture, sub-pulses observed within a given burst can be interpreted as either (1) distinct, shortly-spaced flares which produce a series of individual shocks; or (2) a single shock which generates individual peaks due to complex sub-structure (e.g. emission harmonics) in the maser SED as it drifts downwards across the observing band.\footnote{Drifting behavior can also be imprinted by plasma lensing effects during the propagation to Earth (e.g.~\citealt{Cordes+17,Main+18}) rather than being an intrinsic property of the bursts.  However, such a scenario predicts both downward and upward frequency drifting, in conflict with the always downward drifting seen thus far in FRB 121102 and the CHIME repeaters.}  

In addition to the intrinsic evolution of the SED as the shock propagates outwards, the observed radio flux can be attenuated by inelastic, induced Compton scattering (ICS) of electrons or positrons in the upstream gas  \citep{Lyubarsky08}.  This generates a cut-off in the observed SED below a critical frequency at which the effective optical depth to the induced down-scattering obeys $\tau_{\rm ICS} \gtrsim 1$.  For characteristic parameters, \citetalias{Metzger+19} found that $\tau_{\rm ICS} \gg 1$ near $\nu_{\rm pk}$, in which case the external observer only observes emission at frequencies well above $\nu_{\rm pk}$, at the frequency $\nu_{\rm max} \gg \nu_{\rm pk}$ defined by the condition $\tau_{\rm ICS}(\nu_{\rm max}) \approx 1$.  The downward drift in the observed emission is then controlled by both the evolution of $\nu_{\rm pk}$ and of the condition $\tau_{\rm ICS} \sim 1$.  The derivation of these results is summarized in Appendix \ref{sec:appendix}. 

The standard treatment of ICS assumes that the upstream electrons are subject to weak driving by the electric field $E$ of the FRB wave, i.e.~that the wave-strength (``wiggler'') parameter
\be a \equiv \frac{e E}{2\pi m_{\rm e} c \nu }
\label{eq:strength}
\ee
obeys $ a \ll 1$.  As described in Appendix \ref{sec:appendix_strengthparameter}, our best fit models to the CHIME repeaters imply shocks for which $a \gg 1$, rendering this assumption invalid.  At least for an $e^{-}/e^{+}$ plasma, large driving $a \gg 1$ causes the center of mass motion of the upstream pairs to acquire a high Lorentz factors $\Gamma_{\rm e} \sim a$ away from the shock in the frame of the central engine.  However, once boosting into the co-moving frame of the upstream pairs and accounting for the suppression of the ICS scattering rate $\propto 1/a^{3}$ in the $ a \gg 1$ limit (as found recently by \citealt{Lyubarsky19}), the standard expressions for the ICS optical depth as treated in \citetalias{Metzger+19} assuming a stationary upstream remains unchanged at the order-of-magnitude level (Appendix~\ref{sec:appendix_ICS}).

Given an intrinsic burst duration $\sim t$ at frequency $\nu$ and a measured drift rate $\dot{\nu} \equiv |-d\nu/dt|$,the power-law index of the decay $\nu \propto t^{-\beta}$ can be estimated as
\begin{equation}
\beta \approx \frac{\dot{\nu}t}{\nu} = 0.05\left(\frac{\dot{\nu}}{20\,\rm MHz\,ms^{-1}}\right) \nuCHIME^{-1} t_{-3},
\label{eq:beta}
\end{equation}
where in the second line we have normalized $\nu$ to the center of the CHIME FRB bandpass at 600~MHz, and $t = t_{-3}$~ms.  Inverting eqs.~(45,46) of \citetalias{Metzger+19}, we can relate the measured value of $\beta$ to the power-lax index of the medium $n_{\rm ext} \propto r^{-k}$ according to (eq.~\ref{eq:Appendix_k2})
\begin{align}
k \underset{t \gtrsim \delta t}{=}  
\begin{cases}
\frac{32\beta - 3}{8\beta} ,\,\,\,\,\,\, \beta < \frac{1}{16} ~\leftrightarrow k<-2
\\
\frac{32\beta - 7}{8\beta + 2} ,\,\,\,\,\, \beta>\frac{1}{16} ~\leftrightarrow k>-2
\end{cases}
.
\label{eq:k}
\end{align}
The break in the relationship at $\beta = 1/16$ ($k = -2$) results because for large $k$ the ICS optical depth is dominated by gas on radial scales just outside the shock radius, while for small $k$ the optical depth is dominated by the medium on large radial scales well ahead of the shock.

In general, the mapping $k(\beta)$ will depend on whether the duration of the observed burst $\sim t$ is greater or less than the intrinsic duration of central engine activity, $\delta t$.  Equation (\ref{eq:k}), and other equations in the main text hereafter, are derived under the assumption $t \gtrsim \delta t$.  This is likely to be the relevant regime for a magnetar engine because the observed burst durations $\sim 1-10$~ms exceed the smaller flare duration $\delta t \sim 0.1$ ms (e.g. as set by the Alfv\'en crossing time of a neutron star magnetosphere).  However, it may not be the appropriate limit for other engine models; see Appendix \ref{sec:appendix} for further details and generalization of eq.~\ref{eq:k} to the $t < \delta t$ regime.

\begin{figure}
    \centering
    \includegraphics[width=0.45\textwidth]{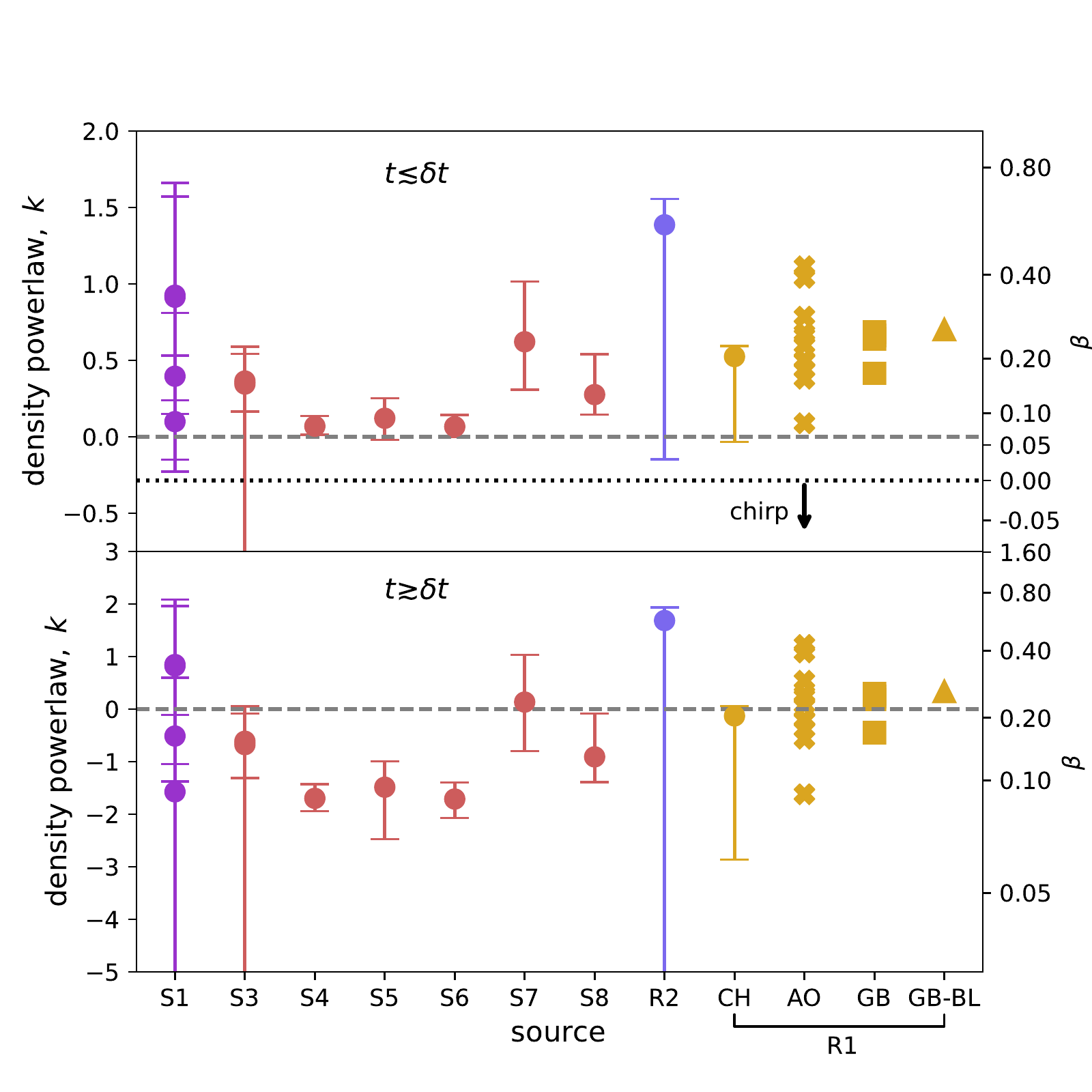}
    \caption{Frequency drift rates for the CHIME repeaters (S1-S8; \citealt{CHIME_newrepeaters}) along with the previously reported CHIME repeater (R2; \citealt{CHIME_R2}) and the original repeater R1 (FRB~121102) as detected by CHIME (CH; \citealt{Josephy+19}) and at higher frequencies by Arecibo Observatory, Green Bank Telescope and the high frequency Breakthrough Listen receiver at Green Bank Observatory (AO, GB, and GB-BL respectively; \citealt{Gajjar+18,Hessels+19}). The right-hand vertical axis shows the values of $\beta$ obtained from CHIME repeaters using equation~(\ref{eq:beta}) while the right-hand axis shows the density index $k$ (such that $n_{\rm ext} \propto r^{-k}$) derived from the \citetalias{Metzger+19} model (eq.~\ref{eq:k}).  Results are shown separately for cases when the observed burst duration, $t$, is less than (top panel) or greater than (bottom panel) the intrinsic duration of the central engine activity, $\delta t$.}
    \label{fig:driftrate}
\end{figure}

Figure~\ref{fig:driftrate} shows the values of $k$ and $\beta$ derived for all repeating FRBs with currently reported drift rates, shown separately assuming $t \gtrsim \delta t$ and $t \lesssim \delta t$.  Across a wide range of sources, and for one specific source (FRB~121102) observed over a wide range in frequency and energy ($\sim0.6-6$ GHz, $\sim 10^{37}-10^{39} \, {\rm erg}$), we find values of $\beta \sim 0.05-0.5$. These imply $k \in [-2,1]$, most of which are in the range of validity of the equations used in \citetalias{Metzger+19} to derive this result.  In particular for $k \gtrsim k_{\rm max} = 4$, the Blandford-McKee solution for blast-wave deceleration used in our analysis is no longer valid (e.g.~\citealt{Blandford&McKee76,Best&Sari00}).\footnote{Though see \citet{Coughlin19} who find evidence supporting validity only up to a shallower value, $k_{\rm max} = 18/7 \simeq 2.57$.}  Also note that for $k < -2/7 \simeq -0.3$ in the $t \lesssim \delta t$ case one predicts $\beta < 0$, i.e. a positive frequency evolution (a ``chirp'' instead of a ``sad trombone''). 

\begin{figure}
    \centering
    \includegraphics[width=0.45\textwidth]{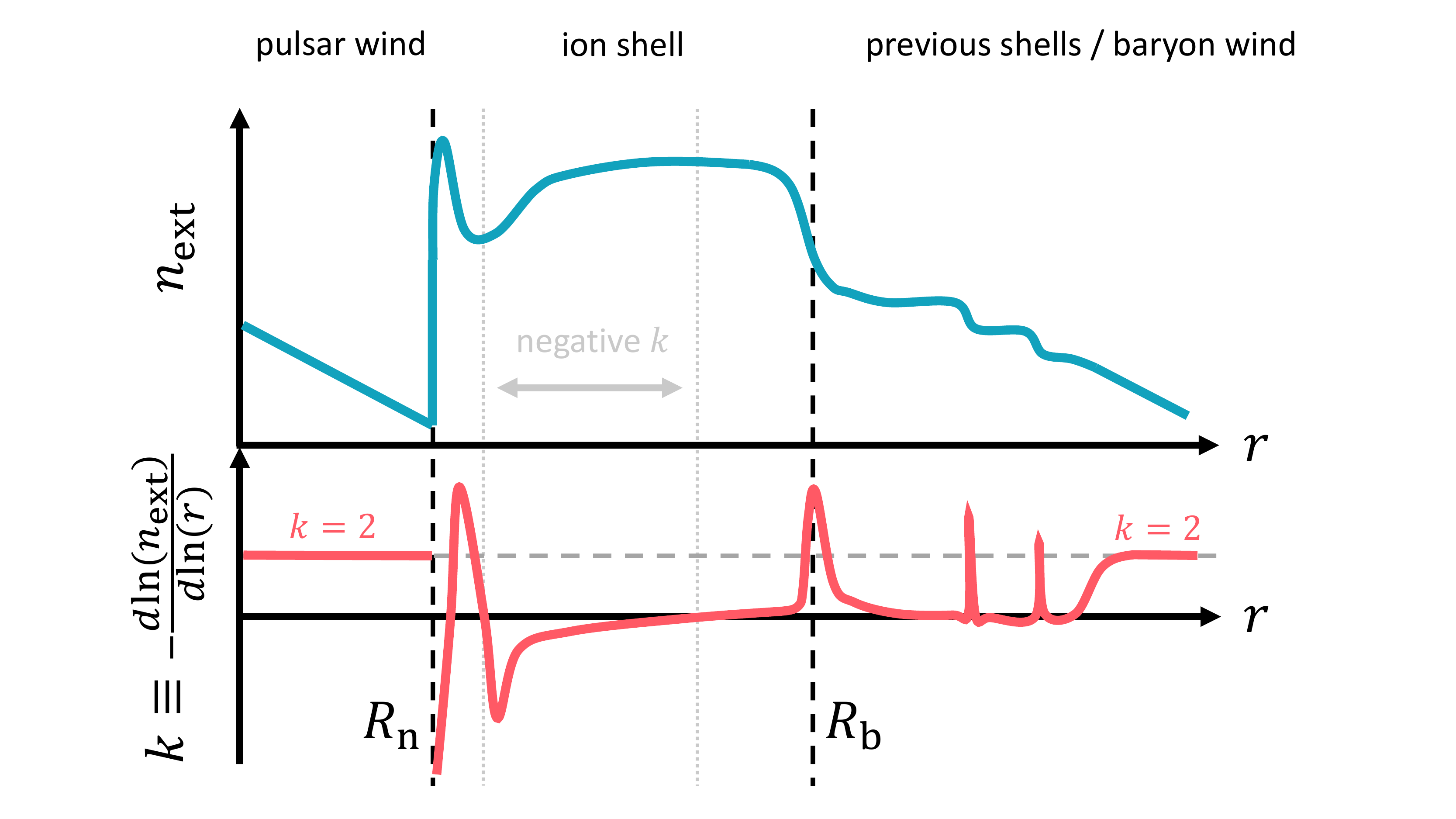}
    \caption{Schematic ambient radial density profile $n_{\rm ext}(r)$ and its power-law slope, 
$k$.
Motivated by modeling of FRB~121102 and the radio afterglow of the giant flare from SGR1806-20, we postulate that the most powerful flares from the central engine (separated by time intervals $\Delta T \sim {\rm d}$) are accompanied by a mildly-relativistic ion-loaded ejecta shell extending to radii $R_{\rm b} = v_w \Delta T \sim 10^{14}-10^{15}$ cm.  At the inner edge of the shell the density profile is rising such that $k < 0$, consistent with our inferences from the drift rate of some bursts (Fig.~\ref{fig:driftrate}).  On large scales $r \gg R_{\rm b}$ the ``train'' of flares ultimately merge into a quasi-steady outflow which may feed a persistent synchrotron nebula.
Between magnetic flares in the case of a magnetar engine, a rotationally-powered wind may inflate a wind bubble behind the baryon shell of radius $R_{\rm n} \ll R_{\rm b}$ (Appendix \ref{sec:appendix_PWN}).    }
    \label{fig:cartoon}
\end{figure}

The values of $k$ that we infer are typically shallower than expected if the upstream medium were that of a steady wind with $n_{\rm ext} \propto r^{-2}$ ($k = 2$), consistent with previous arguments against a wind medium put forward in \citetalias{Metzger+19}.  The flatter $(k \approx 0-1)$ or rising $( k < 0)$ profiles are instead consistent with the ultra-relativistic flare ejecta colliding with the back end of a shell of material released by the previous flare, which has not yet had a chance to become part of a steady flow generated by the accumulation of flares.  Such an interpretation is also consistent with the fact that $\beta$ is observed to vary between bursts from the same repeater.  The latter behavior would be unsurprising if the density profile of the upstream shell were to evolve by the time of the next flare.  Figure \ref{fig:cartoon} shows schematically how the external density profile and that of its power-law index $k$ might appear in such a scenario.

{}

\subsection{Properties of the Flares and External Medium}

Using the observed duration $ t$, fluence $S_\nu$, and frequency $\nu$ as reported for each burst, we again follow the \citetalias{Metzger+19} formalism to solve for the energy and Lorentz factor of the relativistic flare ($E_{\rm flare}$, $\Gamma$) which gives rise to each burst, as well as the value of the external density $n_{\rm ext}$ into which the flare collides as it dissipates its energy. 

Assuming an intrinsic maser SED of shape $\nu L_\nu \propto \nu^{-1}$ at frequencies $\nu > \nu_{\rm pk} \approx 3 \Gamma \nu_{\rm p}$ \citep{Plotnikov&Sironi19}, we find (Appendix \ref{sec:appendix}) 
\begin{align} \label{eq:Eflare}
    E_{\rm flare} \approx 
    2.5 \times 10^{44} \, {\rm erg} \, 
    \left(\frac{m_*}{m_{\rm e}}\right)^{2/5} \left(\frac{f_{\rm e}}{0.5}\right)^{-1/5} 
    f_{\xi,-3}^{-4/5} \nuCHIME^{1/5} \delta t_{-3}^{1/5} \varepsilon_{40}
    ,
\end{align}
where $f_{\xi,-3}$ is the efficiency of the maser normalized to a characteristic value $10^{-3}$ for a cold upstream medium with $\sigma \approx 1$ \citep{Plotnikov&Sironi19} and
\begin{equation} \label{eq:epsilon}
    \varepsilon \equiv 4 \pi \nu S_\nu D^2
\end{equation}
is the isotropic burst energy normalized to $10^{40}$ erg for a source at luminosity distance $D$.  The above expression includes a scaling for the mass $m_*$ of the particle entering the plasma frequency $\nu_{\rm p}$; when the upstream medium is a pair plasma we expect $m_* = m_{\rm e}$, but for an electron-ion plasma the SED could in principle peak lower closer to the ion plasma frequency (due to efficient transfer of energy from ions to electrons ahead of the shock), in which case $m_*/m_{\rm e} = m_{\rm p}/m_{\rm e} = 1836$ for proton composition.  We have also defined the ratio $f_{\rm e} \equiv n_{\rm e} / n_{\rm ext}$ of electron density to the ion density in the upstream medium; values of $f_{\rm e} \simeq 0.5-1$ are expected if the upstream composition is purely electron-ion (\citetalias{Metzger+19}), but in general $f_{\rm e} \gg 1$ if the immediate upstream contains $e^{-}/e^{+}$ pairs (for instance due to a contribution to the upstream medium from the rotationally-powered wind; e.g.~\citealt{Beloborodov19}; see Sect.~\ref{sec:discussion}).  
The expressions presented in this section are only valid for $f_{\rm e} \lesssim m_{\rm p}/m_{\rm e} \approx 10^{3}$, since we assume the swept up ion rest mass dominates the dynamics of blast wave deceleration (though this could easily be generalized).

\begin{figure*}
    \centering
    \includegraphics[width=0.6\textwidth]{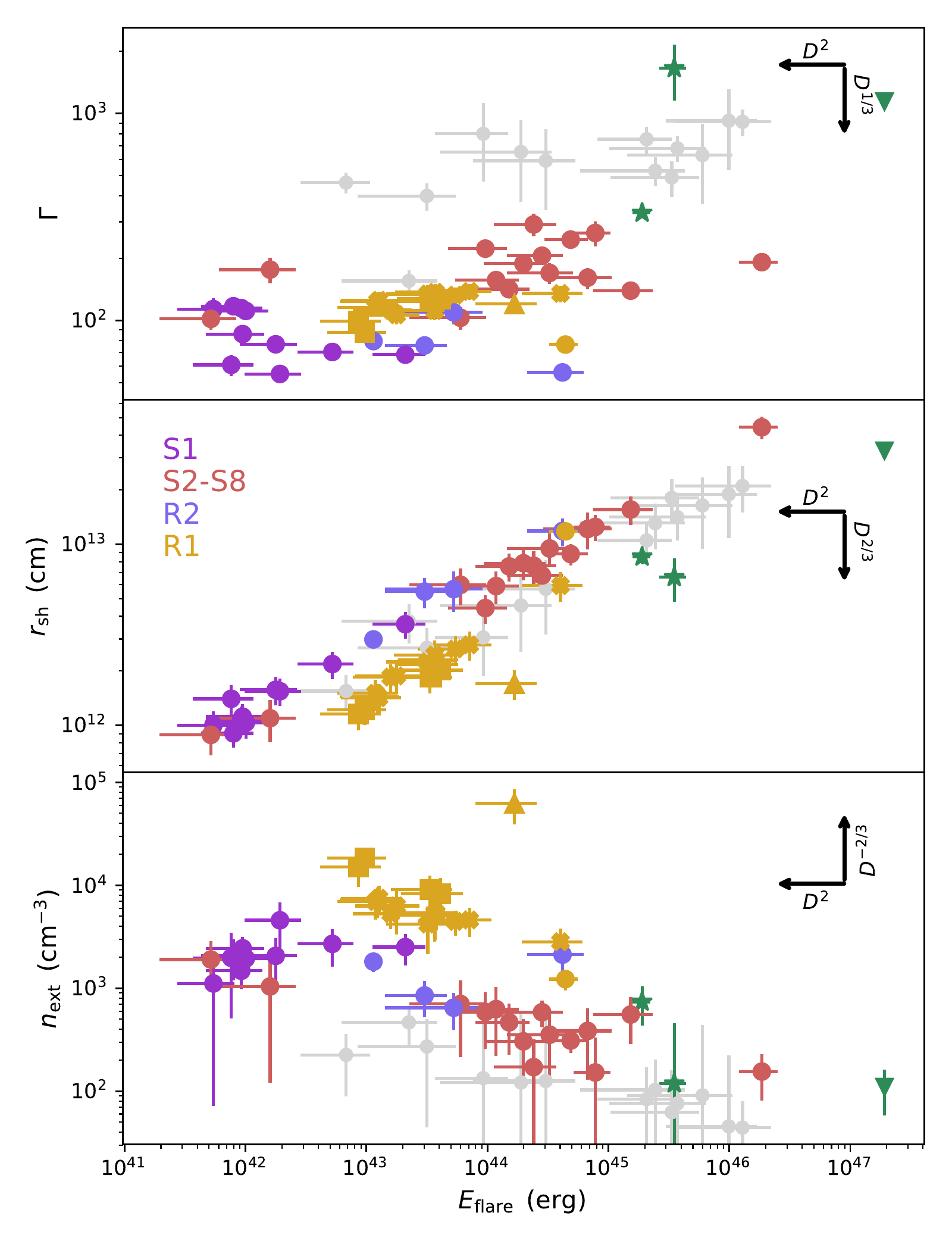}
    \caption{Inferred properties of the FRB-powered shock for the same sample of bursts shown in Fig.~\ref{fig:driftrate}.  As a function of the inferred flare energy $E_{\rm flare}$ we show the Lorentz factor of the shocked gas (top), shock radius $r_{\rm sh}$ (middle), and external density $n_{\rm ext}$ immediately ahead of the shock (bottom).  Green stars and triangle show the same results for the (apparently) non-repeating FRBs with host galaxy candidates identified by ASKAP and DSA-10, respectively (\citealt{Bannister+19,Ravi+19,Prochaska+19}; H. Cho et al. in prep), while grey points show the other published non-repeating CHIME FRBs without host galaxy candidates \citep{CHIME+19c}. 
    We have assumed $m_* = m_{\rm e}$, $f_{\rm e} = 0.5$, $f_{\xi} = 10^{-3}$; scalings with these properties are given by equations (\ref{eq:Gamma}-\ref{eq:next}) in the main text.
    Note that the purple points (`S1') represent the recently localized CHIME repeater, FRB~180916.J0158+65 \citep{Marcote+20}. Along with the yellow and green points, this forms the sub-sample of bursts with known luminosity distances $D$.
    }
    \label{fig:properties}
\end{figure*}

In converting the observed fluence to an energy scale, we require information on the source distances, which except in the case of FRB~121102 \citep{Tendulkar+17} and the recently localized FRBs 180924, 190523, 181112, and 180916.J0158+65 \citep{Bannister+19,Ravi+19,Prochaska+19,Marcote+20} are not yet known or published.  In the following analysis, we estimate the luminosity distance to each source using an approximate redshift-DM relation for the inter-galactic medium, $z \approx {\rm DM}_{\rm E} / 900 \, {\rm pc \, cm}^{-3}$ \citep{Zhang18}, where  ${\rm DM}_{\rm E} = {\rm DM} - {\rm DM}_{\rm MW}$ is the extra-Galactic component of the dispersion measure.
However, note that significant variation in this mean relation is expected along different lines of sight \citep{McQuinn14,Prochaska&Zheng19}.
We adopt ${\rm DM}_{\rm MW} = 200 \, {\rm pc \, cm}^{-3}$ and $80 \, {\rm pc \, cm}^{-3}$ for sources S1 and S2, respectively \citep{CHIME_newrepeaters}, ${\rm DM}_{\rm MW} = 120 \, {\rm pc \, cm}^{-3}$ for R2 \citep{CHIME_R2}, and conservatively take ${\rm DM}_{\rm MW} = 0 \, {\rm pc \, cm}^{-3}$ for S3--S8 for which no Galactic DM estimate could be found. For FRB~121102 we use the known redshift and luminosity distance, $z \simeq 0.193$ and $D \simeq 972 \, {\rm Mpc}$ \citep{Tendulkar+17}.  Since our distance estimate neglects any DM contributions in the local source vicinity or host galaxy (and for S3-S8 even further neglects any Galactic DM), it should be interpreted as a loose upper-limit on the true source distances. The scaling of inferred source properties with $D$ is easily given by re-scaling $\varepsilon \propto D^2$ in each equation, as indicated explicitly in equation~(\ref{eq:Appendix_zD_scaling}) and Fig.~\ref{fig:properties}.

The bulk Lorentz factor of the shocked gas behind the blast-wave at the epoch of FRB emission can be written as
\begin{eqnarray} \label{eq:Gamma}
    \Gamma \approx
    324 \left(\frac{m_*}{m_{\rm e}}\right)^{1/30} \left(\frac{f_{\rm e}}{0.5}\right)^{1/15} f_{\xi,-3}^{-1/15} \nuCHIME^{-7/30}  t_{-3}^{-2/5} \varepsilon_{40}^{1/6} .
 \end{eqnarray}
From equation~(\ref{eq:Gamma}), the shock radius at the emission epoch of the observed FRB is given by
\begin{eqnarray}
    r_{\rm sh} \simeq 2 \Gamma^2 c t 
    \approx
    6 \times 10^{12} \, {\rm cm} \, 
    \left(\frac{m_*}{m_{\rm e}}\right)^{1/15} \left(\frac{f_{\rm e}}{0.5}\right)^{2/15} 
    \\ \nonumber
    \times
    f_{\xi,-3}^{-2/15} \nuCHIME^{-7/15}  t_{-3}^{1/5} \varepsilon_{40}^{1/3}.
\end{eqnarray}
Finally, the density of the external medium at $r_{\rm sh}$ is given by 
\begin{eqnarray}
    n_{\rm ext}(r_{\rm sh}) \approx
    248 \, {\rm cm}^{-3} \,
    \left(\frac{m_*}{m_{\rm e}}\right)^{2/15}\left(\frac{f_{\rm e}}{0.5}\right)^{-11/15} 
    \\ \nonumber
    \times
    f_{\xi,-3}^{-4/15} \nuCHIME^{31/30} t_{-3}^{2/5} \varepsilon_{40}^{-1/3} 
\label{eq:next}
\end{eqnarray}
We neglect redshift effects on the observed burst frequency and duration since the DM-inferred source redshifts are only loose upper limits.  Equation~(\ref{eq:Appendix_zD_scaling}) provides the scaling with redshift of the derived quantities, which show these are only minor corrections. 
Equation~(\ref{eq:next}) can also be re-expressed in terms of the upstream electron (positron) density,
\begin{eqnarray} 
    &&n_{\rm e}(r_{\rm sh})
    = f_{\rm e} n_{\rm ext}(r_{\rm sh})
     \\ \nonumber
    &\approx&
    124 \, {\rm cm}^{-3} \,
    \left(\frac{m_*}{m_{\rm e}}\right)^{2/15}\left(\frac{f_{\rm e}}{0.5}\right)^{4/15} 
    f_{\xi,-3}^{-4/15} \nuCHIME^{31/30} t_{-3}^{2/5} \varepsilon_{40}^{-1/3} .
\end{eqnarray}

\begin{figure}
    \centering
    \includegraphics[width=0.45\textwidth]{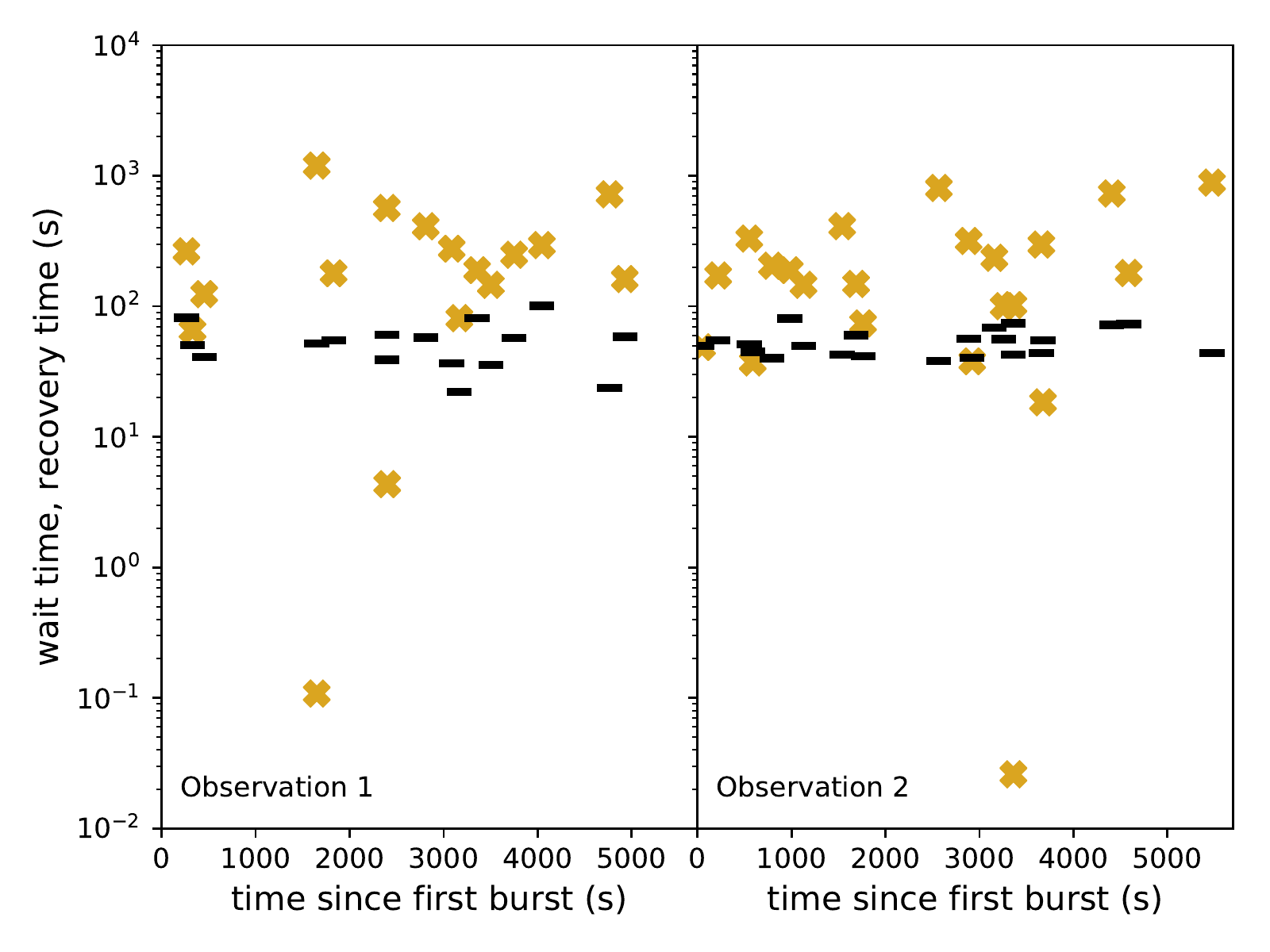}
    \caption{Dynamical timescale, $t_{\rm dyn} \sim r_{\rm sh}/c$, for recovery of the shocked gas (black bars) compared to the wait time until the next burst, $t_{\rm wait}$ (yellow crosses), for the sample of bursts from FRB~121102 of \citet{Gourdji+19}, divided into the two observation epochs of that sample. In nearly all cases $t_{\rm wait} \gtrsim t_{\rm dyn}$ consistent with the assumption of the synchrotron maser blast-wave model of a cold upstream medium.}
    \label{fig:waittime}
\end{figure}

Figure~\ref{fig:properties} shows the values of $\Gamma$, $r_{\rm sh}$, and $n_{\rm ext}(r_{\rm sh})$ as a function of $E_{\rm flare}$ for all the bursts from our repeating sample, under the fiducial assumptions $f_{\xi} = 10^{-3}$, $m_* = m_{\rm e}$, and $f_{\rm e} = 0.5$.  Shown for comparison with green points are the same results for the (apparently) non-repeating FRBs with host galaxy candidates identified by ASKAP and DSA-10 (\citealt{Bannister+19,Ravi+19,Prochaska+19}; H. Cho et al. in prep), while grey points show the other published non-repeating CHIME FRBs without host galaxy candidates \citep{CHIME+19c}.  The flare energies we derive range from $\sim 10^{43}-10^{46}$ erg, depending roughly inversely on efficiency $f_{\xi}$.  Our results are relatively insensitive to the values of $f_{\rm e}$ and $m_*$, with the exception of $n_{\rm ext} \propto f_{\rm e}^{-11/15}$ which could be significantly reduced by an upstream medium heavily loaded in $e^{-}/e^{+}$ pairs ($f_{\rm e} \gg 1$; although even in this case, the electron/positron density $n_{\rm e} \propto f_{\rm e}^{4/15}$ is only mildly affected).  

The values of the Lorentz factors that we infer, $\Gamma \sim 10^2-10^3$, are actually lower limits on the initial Lorentz factors of the flare ejecta, since in general the blast-wave will have decelerated from its initial speed by the time the FRB enters the instrumental bandpass.  Also worth noting is that the shock radii we infer, $r_{\rm sh} \sim 10^{12}-10^{13}$ cm, are safely smaller than the mean radii $R_{\rm b} \approx v_{\rm w} \Delta T \sim 10^{14}-10^{15}$ cm of a baryon ejecta shell expanding at $v_{\rm w} \lesssim c$ released from a previous major flare each interval $\Delta T \sim 10^{4}-10^{5}$ s (the rare, powerful flares \citetalias{Metzger+19} hypothesize create the upstream medium; Fig.~\ref{fig:cartoon}).  

The inferred upstream densities can also be compared to those needed on radial scales $\sim R_{\rm b}$ to generate a given baryon mass injection rate $\dot{M}$ into the engine environment on large scales, 
\begin{align}
\dot{M} \sim \frac{m_{\rm p} n_{\rm sh}R_{\rm b}^{3}}{\Delta T} 
\approx  6 \times 10^{20}{\rm g\,s^{-1}}\left(\frac{n_{\rm sh}}{10^{4}\,{\rm cm^{-3}}}\right)\left(\frac{\Delta T}{10^{5}{\rm s}}\right)^{2}\left(\frac{v_w}{0.5c}\right)^{3}.
\label{eq:mdot}
\end{align}
The values $n_{\rm sh} \sim 10^{3}-10^{4}$ cm$^{-3}$ we infer for the repeating FRB sample imply values $\dot{M} \sim 10^{20}-10^{21}$ g s$^{-1}$, compatible with the mass injection rate  \citet{Margalit&Metzger18} found was needed to explain the persistent synchrotron flux and RM of FRB~1211012 as being generated in a flare-powered nebula.

As a final consistency check, the upstream densities and shock radii can be converted into minimum DM contributions to the FRB emission, ${\rm DM}_{\rm ext} \approx f_{\rm e} n_{\rm ext} r_{\rm sh}$.  These values are typically $\sim 10^{-4}-10^{-2}$ pc cm$^{-3}$, orders of magnitude smaller than the total burst DM. 
However, if a medium of similar or greater density (as supported by our inference of $k \le 0$; Fig.~\ref{fig:driftrate}) were to persist to the much larger radii $R_{\rm b} \gtrsim 100r_{\rm sh}$ of the hypothesized ion shell, then the shell's total DM contribution would be larger by at least a couple orders of magnitude.  Evolution of the gas column through the expanding ion shell(s) could then result in measurable stochastic variations in the DM on timescales $\Delta T \sim$ days between major flares \citepalias{Metzger+19}.

Efficient synchrotron maser emission requires the electrons in the upstream medium to be ``cold'', i.e. if their velocities are isotropically distributed that they possess a sub-relativistic temperature $kT \ll m_{\rm e} c^{2}$ (Babul \& Sironi, in prep).  However, even an initially cold upstream will be heated by the passage of the FRB-generating shock, raising the temperature of the medium into which the next flare would collide.  This might act to suppress FRB emission after a given burst on timescales shorter than the dynamical time, $t_{\rm dyn} \sim r_{\rm sh}/c$, at the shock radius, over which the upstream medium will have a chance to ``recover'' by expanding and cooling adiabatically.  Figure~\ref{fig:waittime} shows our derived values of $t_{\rm dyn}$ in comparison to the measured interval since the last burst, $t_{\rm wait}$, from the \citet{Gourdji+19} burst sample from FRB~121102.  We see that in most cases $t_{\rm dyn} > t_{\rm wait}$, providing another consistency check on our general scenario.  Indeed, we are led to speculate that the typical absence of bursts with shorter wait times $t_{\rm wait} \lesssim t_{\rm dyn}$ might not indicate a complete lack of flares/shocks on such short timescale, but suppression of the FRB emission due to the upstream being too hot when flares are spaced too closely together.  

\subsection{Rates and Energetics}

The bursts from FRB~121102 exhibit a complex repetition pattern with a non-Poissonian distribution and a median interval of hundreds of seconds (e.g.~\citealt{Oppermann+18,Katz18,Li+19,Gourdji+19}).  
The luminosity function of FRB~121102 may also vary in time.  \citet{Law+17} find that, weighted by their radiated energy (and thus the flare energy in our scenario), the luminosity function is dominated by the rare highest fluence bursts, which take place at a rate $\mathcal{R} \lesssim 1$ per day \citep[see also][]{Nicholl+17}, while observations of low-energy bursts at different epochs imply a steeper luminosity function whose total energy is dominated by frequent less-energetic bursts \citep{Gourdji+19}.

\begin{figure}
    \centering
    \includegraphics[width=0.45\textwidth]{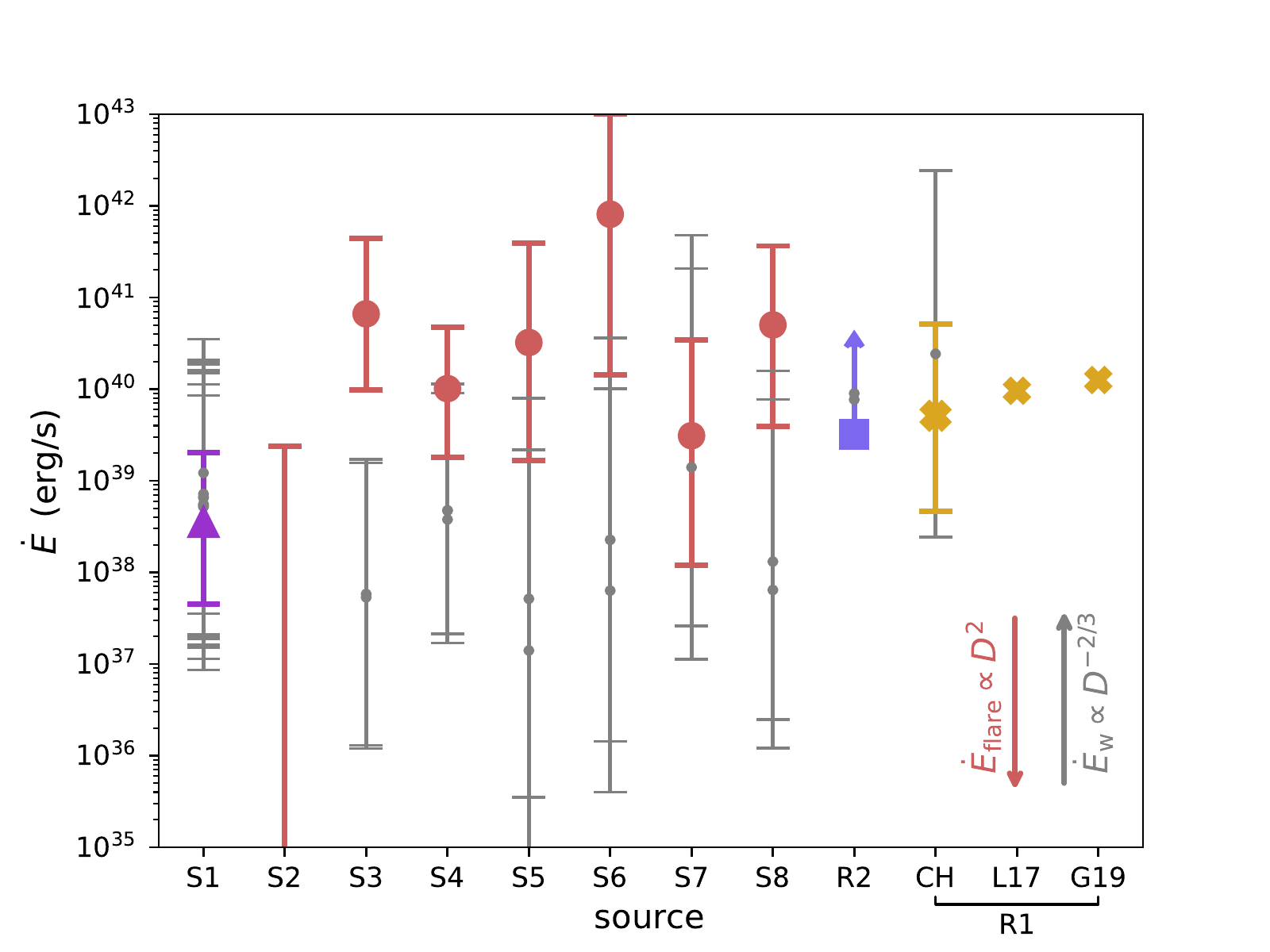}
    \caption{Inferred engine power emitted in the relativistic flares that produce the bursts, $\dot{E}_{\rm flare}$ (eq.~\ref{eq:Edot_flare}; color symbols), in comparison with the power emitted carried away by slower-moving baryon-loaded shells into which the relativistic flares collide, $\dot{E}_{\rm w}$ (eq.~\ref{eq:Edot_w}; grey points). Symbols and notation are as in Fig.~\ref{fig:driftrate}, with the addition of L17 and G19 for rate and energetic constraints on R1 (\citealt{Law+17} and \citealt{Gourdji+19}, respectively). The implied power in both outflows is comparable within the large uncertainties, especially considering that $\dot{E}_{\rm flare}$ should be decrease and $\dot{E}_{\rm w}$ increase if the source distances are smaller than our conservative estimate (scaling shown with arrows). }
    \label{fig:Edot}
\end{figure}

Combining our results from the previous section for the average flare energy, $\langle E_{\rm flare} \rangle$, with the measured repetition rates of the CHIME-detected repeaters \citep{CHIME_R2,CHIME_newrepeaters}, and assuming the latter are among the more energetic flares, we may crudely estimate the flare energy injection rate by the central engine for each repeating source as
\begin{equation} \label{eq:Edot_flare}
    \dot{E}_{\rm flare} \sim \left\langle E_{\rm flare} \right\rangle \mathcal{R} 
    \approx 3 \times 10^{39} {\rm erg \, s}^{-1} \, \left\langle E_{\rm flare} \right\rangle_{44} \mathcal{R}_{0.1{\rm hr}^{-1}} ,
\end{equation}
where $\mathcal{R}_{0.1{\rm hr}^{-1}}$ is the repetition rate normalized to once per ten hours and $\left\langle E_{\rm flare} \right\rangle_{44}$ the average flare energy normalized to $10^{44}$~erg.  
Although the uncertainties are large, Figure~\ref{fig:Edot} shows typical values $\dot{E}_{\rm flare} \sim 10^{40}-10^{41}$ erg s$^{-1}$ for the repeater population.

What should this energy injection rate be compared to?  If the upstream medium setting $n_{\rm ext}$ is a baryon-loaded trans-relativistic shell ejected a time $\Delta T \sim \mathcal{R}^{-1}$ before the burst, then the total mass in such a shell is $\propto n_{\rm sh}R_{\rm b}^{3} \sim n_{\rm ext} v_{\rm w}^3 \Delta T^3$  (see eq.~\ref{eq:mdot} and surrounding discussion). This then allows an estimate of the kinetic wind luminosity required to generate the baryon-loaded medium into which the relativistic flares collide,
\begin{align} \label{eq:Edot_w}
    \dot{E}_{\rm w} &\sim \frac{2\pi}{3} n_{\rm ext} m_{\rm p} v_{\rm w}^5 \mathcal{R}^{-2} \nonumber \\
    &\approx 3 \times 10^{40} \, {\rm erg \, s}^{-1} \, \left(\frac{v_{\rm w}}{0.5 c}\right)^5\left(\frac{n_{\rm ext}}{10^{4}{\rm cm^{-3}}}\right) \mathcal{R}_{0.1{\rm hr}^{-1}}^{-2} ,
\end{align}
where the ion shell velocity is normalized to a value $v_{\rm w} \approx (0.5-0.8)c$ similar to that needed to explain the RM and persistent radio emission from FRB~121102 \citep{Margalit&Metzger18}.  Observations of the radio afterglow of the Galactic giant flare from SGR 1806-20 also support the ejection of mildly relativistic material during the flare ($v_{\rm w} \approx 0.7$ c; \citealt{Gaensler+05,Gelfand+05,Granot+06}).  

Figure~\ref{fig:Edot} shows that the allowed values of $\dot{E}_{\rm flare}$ and $\dot{E}_{\rm w}$ (calculated for $v_{\rm w} = 0.5 \,c$) typically overlap, albeit within the several order-of-magnitude uncertainties of each quantity. Though not a requirement of our model, it would also not be surprising if the energy released promptly from the engine (as ultra-relativistic ejecta during the earliest stages of the flare) were to be comparable to that released over longer timescales in slower baryon-loaded outflow during the aftermath of the flare.  For the fiducial parameters required to fit the persistent source luminosity, rotation measure, and burst properties of FRB~121102, these energies were also found to be comparable \citep{Margalit&Metzger18}.  It is thus non-trivial that the CHIME repeaters also show tentative evidence for equality between $\dot{E}_{\rm flare}$ and $\dot{E}_{\rm w}$ given the dependence of such agreement on the burst repetition rate $\mathcal{R}$, which is otherwise independent of the other observed burst properties (frequency, duration, luminosity).

Our above analysis assumes that the relativistic ejecta from the central engine is released isotropically.  If the ejecta were instead beamed into a limited fractional solid angle $f_{\rm b} < 1$, then the true (beaming-corrected) energy of a given flare $E_{\rm flare}$ would decrease by a factor $f_{\rm b}$.  On the other hand, the same beaming would cause one to underestimate the flaring rate by the same factor $\mathcal{R} \propto f_{\rm b}^{-1}$, such that $\dot{E}_{\rm flare} \propto E_{\rm flare}\mathcal{R}$ would remain unchanged from that inferred from the isotropic case provided that the flares are emitted into random directions.  The angular distribution of the flare direction, and indeed whether beaming is relevant at all, depends on the engine model.  A neutron star, for instance, may possess a preferred direction defined by its magnetic dipole axis.  However, as long as the rotational period of the star exceeds the burst duration $\delta t$ then flares which occur at random rotational phases would generate a flare distribution which is at least azimuthally-symmetric.  

The same shock responsible for generating coherent synchrotron maser emission (the FRB) also produces an incoherent synchrotron afterglow, analogous to a scaled-down version of those which accompany the interaction of gamma-ray burst (GRB) jets with the circum-burst environment (\citealt{Lyubarsky14}, \citetalias{Metzger+19}, \citealt{Beloborodov19}).  However, unlike normal GRB afterglows the emission is produced by {\it thermal} electrons heated at the shock rather than a power-law non-thermal distribution (e.g.~\citealt{Giannios&Spitkovsky09,Sironi&Spitkovsky09}).  The shock-heated thermal electrons are fast-cooling in the post-shock magnetic field in the case of an upstream electron-ion medium for typical FRB parameters (\citetalias{Metzger+19}), in which case an order-unity fraction of the blast wave energy will be radiated (in contrast to the much smaller fraction $f_{\xi} \lesssim 10^{-3}$ radiated as an FRB).  

The peak frequency of the synchrotron afterglow at times $t \gtrsim \delta t$ is given by (\citetalias{Metzger+19}, their eq.~57)
\be
h\nu_{\rm syn} \approx 228\,{\rm MeV} \, \left(\frac{f_{\rm e}}{0.5}\right)^{-2} \sigma_{-2}^{1/2} E_{\rm flare,44}^{1/2} t_{-3}^{-3/2},
\label{eq:nusyn}
\ee
where $\sigma_{-2} \equiv \sigma/10^{-2}$.  For an electron-ion upstream medium ($f_{\rm e} \sim 1$) the emission should occur in the MeV or GeV gamma-ray band, while for a pair-loaded upstream medium ($f_{\rm e} \gg 1$) the afterglow would peak in the X-ray or optical band (\citetalias{Metzger+19}; \citealt{Beloborodov19}).  

\begin{figure}
    \centering
    \includegraphics[width=0.45\textwidth]{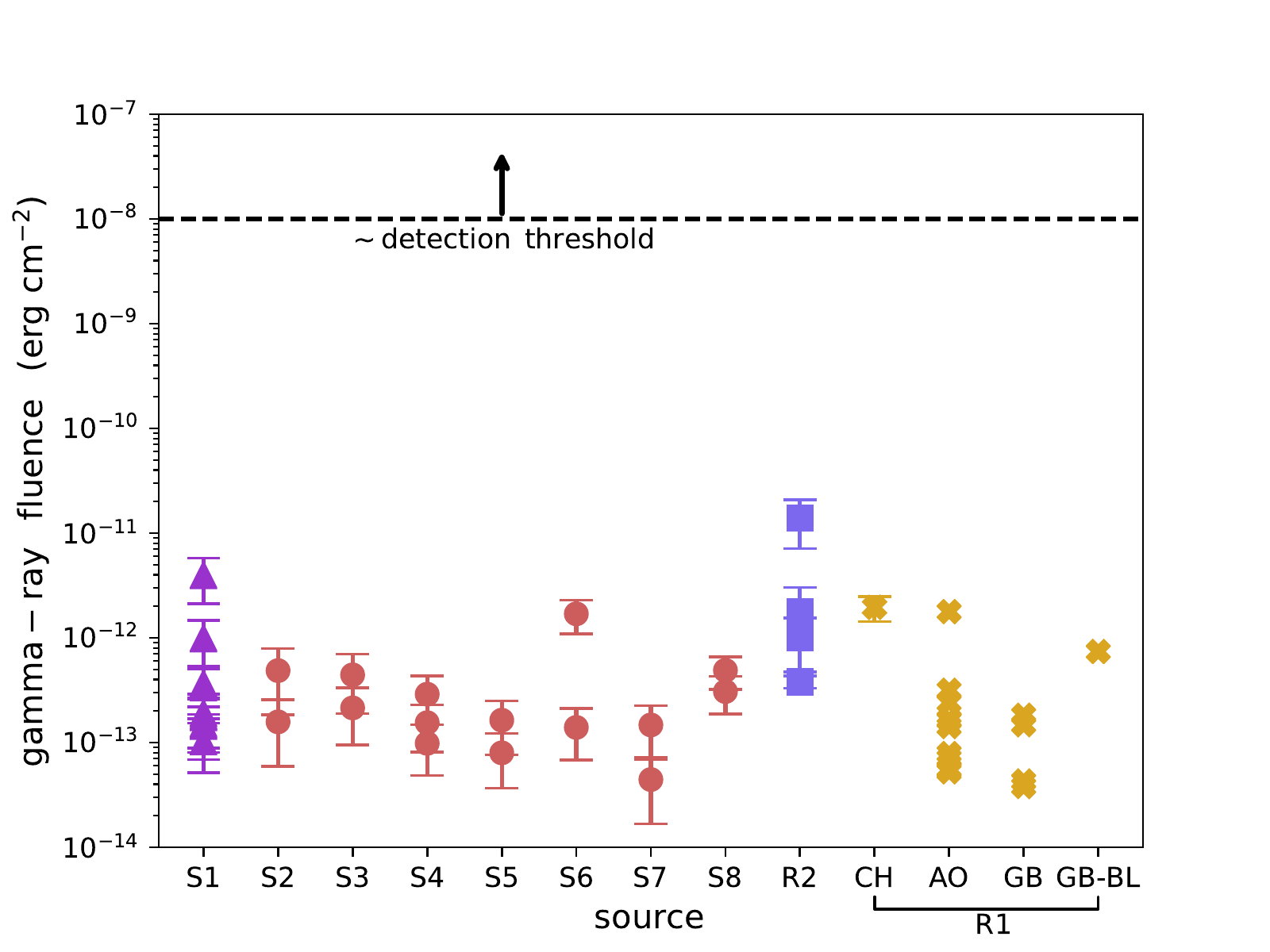}
    \caption{Predicted maximum fluence of thermal gamma-ray afterglow associated with each repeating FRB source (\citetalias{Metzger+19}). The predicted fluences are at least $3-4$ orders of magnitudes lower than the lowest fluence GRBs ever detected $\sim 10^{-8}$ erg cm$^{-2}$. The symbol, color, and naming conventions follow Fig.~\ref{fig:driftrate}.}
    \label{fig:GRB}
\end{figure}

Under the most optimistic assumption that electrons are heated to rough equipartition with the ions, each FRB is accompanied by a gamma-ray burst of duration $\sim \delta t$ and maximum fluence $E_{\gamma} \approx 0.5E_{\rm flare}$.  Figure~\ref{fig:GRB} shows the predicted maximum gamma-ray fluences of the bursts from the repeating FRBs in our sample.  Typical values are $F \sim 10^{-12}$~erg~cm$^{-2}$, several orders of magnitude lower than the lowest fluence GRB detected by {\it Fermi} Gamma-ray Burst Monitor ($\sim 10^{-8}$~erg~cm$^{-2}$; \citealt{vonKienlin+14}).  It is thus unsurprising that no convincing gamma-ray counterparts to FRBs have yet been detected (e.g.~\citealt{Bannister+12,Palaniswamy+14,DeLaunay+16,Cunningham+19}).  

Prospects are better to detect the afterglow in the visual band, but only if the upstream medium of the shock is much higher density (e.g.~in dark FRB-free phases right after baryon ejection from major flares) or if the upstream medium is sufficiently pair-loaded ($f_{\rm e} \gg 10^{2}-10^{3}$) for $h\nu_{\rm syn}$ to fall in the UVOIR band.  The best current upper limits on optical radiation simultaneous with a burst from FRB 121102 are $\lesssim 10^{47}$ erg s$^{-1}$ on timescales $\lesssim 70$ ms.  Future instruments such as HiPERCAM \citep{Dhillon+16} or the planned Ultra-Fast Astronomy observatory \citep{Li+19} will be more sensitive by several orders of magnitude and thus could place interesting limits.

\section{Discussion}
\label{sec:discussion}

\subsection{Challenges and Uncertainties of the Model}
\label{sec:variations}

The main results of the previous section were derived within the nominal assumptions of the \citetalias{Metzger+19} model, particularly that the upstream medium --- at least on sufficiently large scales --- is baryon-loaded and expanding at mildly relativistic speeds \citep{Margalit&Metzger18}.
However, some potential concerns exist about the upstream medium being a slowly-expanding electron-ion plasma (e.g.~as raised by \citealt{Beloborodov19}), which we address in this section.  We also consider to what extent our results would change, and these concerns alleviated or worsened, if the upstream behind the ion shell were instead a nebula dominated by $e^{-}/e^{+}$ pairs ($f_{\rm e} \gg 1$), e.g. from a pulsar wind (see also $\S\ref{sec:upstream}$).

The fiducial value for the maser efficiency adopted in our analysis, $f_{\xi} \sim 10^{-3}$, is calibrated off of PIC simulations performed for an $e^{-}/e^{+}$ upstream medium and its value could in principle be smaller in the electron-ion case by a factor of up to $\sim m_{\rm e}/m_{\rm ion} \sim 10^{-3}-10^{-4}$, corresponding to the small fraction of the upstream kinetic energy carried by the electrons (which ultimately power the synchrotron maser).  In such a case, the required energy budget of the flares $E_{\rm flare} \propto f_{\xi}^{-4/5}$ for an electron-ion upstream would increase by several orders of magnitude, and possibly beyond plausible values for some engine models.  Contrary to this expectation, \citet{Iwamoto+19} present 2D PIC simulations showing that the kinetic energy of the electrons entering the shock is enhanced by the wakefield electric field created in the precursor region (as electrons are separated away from the ions by the ponderomotive force of the precursor, see e.g., \citealt{Hoshino08}), such that the effective efficiency is close to the upstream pair case.  However, final conclusions about the maser efficiency in the electron-ion case awaits future multi-dimensional PIC simulations which run sufficiently long to achieve a steady-state in the precursor emission.

Another concern in invoking an ion-dominated upstream is whether the magnetization $\sigma \propto B^{2}/\rho$, into which the relativistic flares are colliding, is too low for efficient maser emission.  Consider that if the baryon shell is ejected in a quasi-steady outflow of magnetization $\sigma = \sigma_0 \sim 1$ from the central engine over a time $T_{\rm ion}$, then since $B \propto 1/r$ and $\rho \propto 1/r^{2}$ in a steady wind, we have $\sigma \sim {\rm constant} \sim \sigma_0$ out to radii $\Delta = v_{\rm w} T_{\rm ion} \sim 10^{10}(T_{\rm ion}/1{\rm s})(v_{\rm w}/0.5c)$~cm.  At radii $r \gg \Delta$, the ejecta shell would soon achieve homologous expansion, for which $\rho \propto 1/r^{3}$ and $B \propto 1/r^{2}$, such that $\sigma \propto 1/r$ could in principle decrease by several orders of magnitude between $r \sim \Delta$ and the radii $r_{\rm sh} \sim 10^{12}-10^{13}$~cm of the inferred shocks (Fig.~\ref{fig:properties}).  For instance, in the case of an ion shell released from a magnetar giant flare (e.g.~$T_{\rm ion} \sim 0.1-1$ s, similar to the observed durations of Galactic giant flares) we could expect $\sigma(r_{\rm sh})\sim 10^{-2}-10^{-3}$ for $\sigma_0 \sim 1$ if the upstream medium is the ion shell.  

A low value of $\sigma$ is problematic for the maser emission because the downstream electrons should gyrate about the ordered magnetic field, which could be disrupted by the Weibel instability which occurs for sufficient low values of $\sigma$ \citep[e.g,.][]{sironi_13}. In the absence of efficient transfer of energy from ions to electrons ahead of the shock (e.g., as due to the wakefield discussed before), electrons would enter the shock with an effective magnetization $\sigma_{\rm e}$, which is a factor of $m_{\rm ion}/m_{\rm e} \gtrsim 10^{3}$ larger than the normal $\sigma$ for the ions defined above. For values of $\sigma(r_{\rm sh}) \gtrsim 10^{-3}$ in the ion shell (as motivated above), electrons would still have $\sigma_{\rm e} \gtrsim 1$, which is sufficiently large (i.e., the electrons are sufficiently magnetized) for producing the synchrotron maser. However, such a scenario would suffer from the poor efficiency mentioned above, given the small fraction of the upstream kinetic energy carried by the electrons. Whether the shock at moderately low magnetization effectively self-regulates to give efficient precursor emission despite the adverse effect of the Weibel instability would require long-term multi-dimensional PIC simulations. For relatively short run times, \citet{Iwamoto+19} have shown that in 2D PIC simulations of electron-ion shocks with mass ratio of 50, a relatively low value of  $\sigma\sim 0.02$ can still produce precursor waves with $f_\xi\sim 10^{-3}$ (with the only caveat that the shock was yet to reach a steady state in the precursor emission).

\begin{figure}
    \centering
    \includegraphics[width=0.45\textwidth]{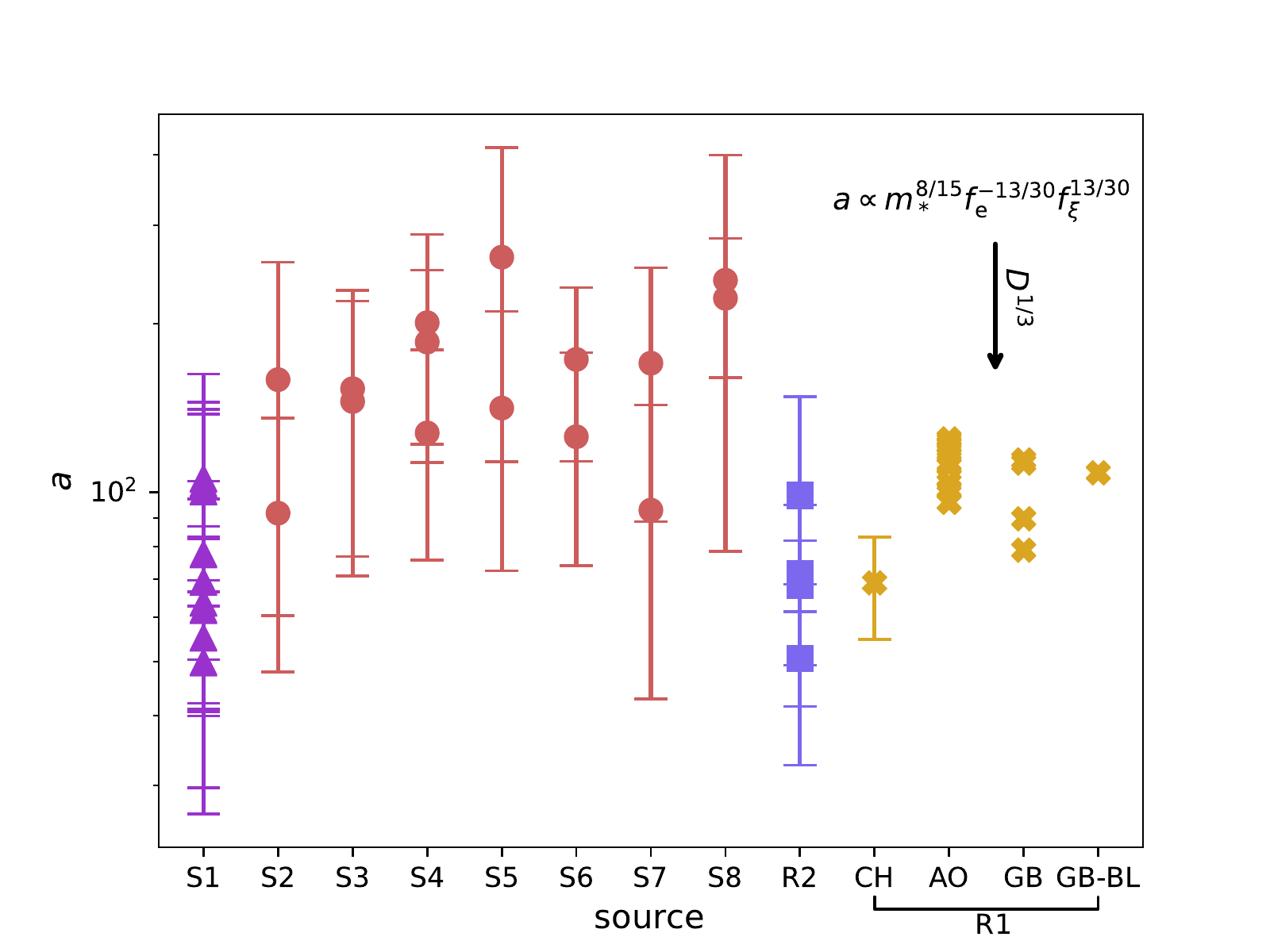}
    \caption{The inferred wave-strength parameter $a$ (eq.~\ref{eq:Appendix_a0}) for FRBs in our sample (color and naming convention follow Fig.~\ref{fig:driftrate}).  We have assumed $m_* = m_{\rm e}$, $f_{\rm e} = 0.5$, $f_{\xi} = 10^{-3}$.}
    \label{fig:strengthparameter}
\end{figure}

Another uncertain issue is the effect of the strong precursor wave on the properties of the upstream medium and the efficiency of the maser.  Using the shock properties of our FRB sample presented in $\S\ref{sec:model}$, we infer wave-strength parameters $a \sim 100$ at the time of the observed FRBs under the standard assumption of an electron-ion upstream (see Fig.~\ref{fig:strengthparameter}).  The anticipated effect of an ultra-strong wave $a \gg 1$ is to induce bulk motion of the pair gyrocenter motions along the direction of the wave propagation away from the shock at a Lorentz factor $\Gamma_{\rm e} \sim a$ (e.g.~\citealt{Arons72,Blandford72,lyubarsky_06,Lyubarsky19}).  

Following the numerical set-up outlined in \citet{Plotnikov&Sironi19}, we have performed a suite of one-dimensional PIC simulations of ultra-relativistic shocks propagating into an upstream pair plasma of magnetization $\sigma = 3$, in order to explore the effects of large strength parameter $a \gg 1$ on the dynamics of the upstream medium and the properties of the maser emission.  Our results are shown in Figure \ref{fig:PIC} and described in greater detail in Appendix~\ref{sec:appendix_strengthparameter}.

Firstly, we confirm that for $a \gg 1$ the precursor wave acts to accelerate (``push'') upstream electron fluid away from the shock to a bulk Lorentz factor $\sim a$ and to induce relativistic dispersion in the particle momentum, consistent with theoretical expectations (e.g.~\citealt{Lyubarsky19}).  In principle, such pre-acceleration and heating of the upstream could change the rate of induced Compton scattering, which controls the observed peak of the SED in the \citetalias{Metzger+19} model.  However, as shown in Appendix \ref{sec:appendix_ICS}, the expected suppression of the ICS rate in the co-moving frame of the pair gyrocenter \citep{Lyubarsky19}, coupled with appropriate Lorentz transformation of the in-beam and out-of-beam scattering rates, results in the ICS optical depth remaining unchanged in the $a \gg 1$ limit from what one would infer for a stationary upstream medium of the same electron column.

\begin{figure*}
    \centering
    \includegraphics[width=0.6\textwidth]{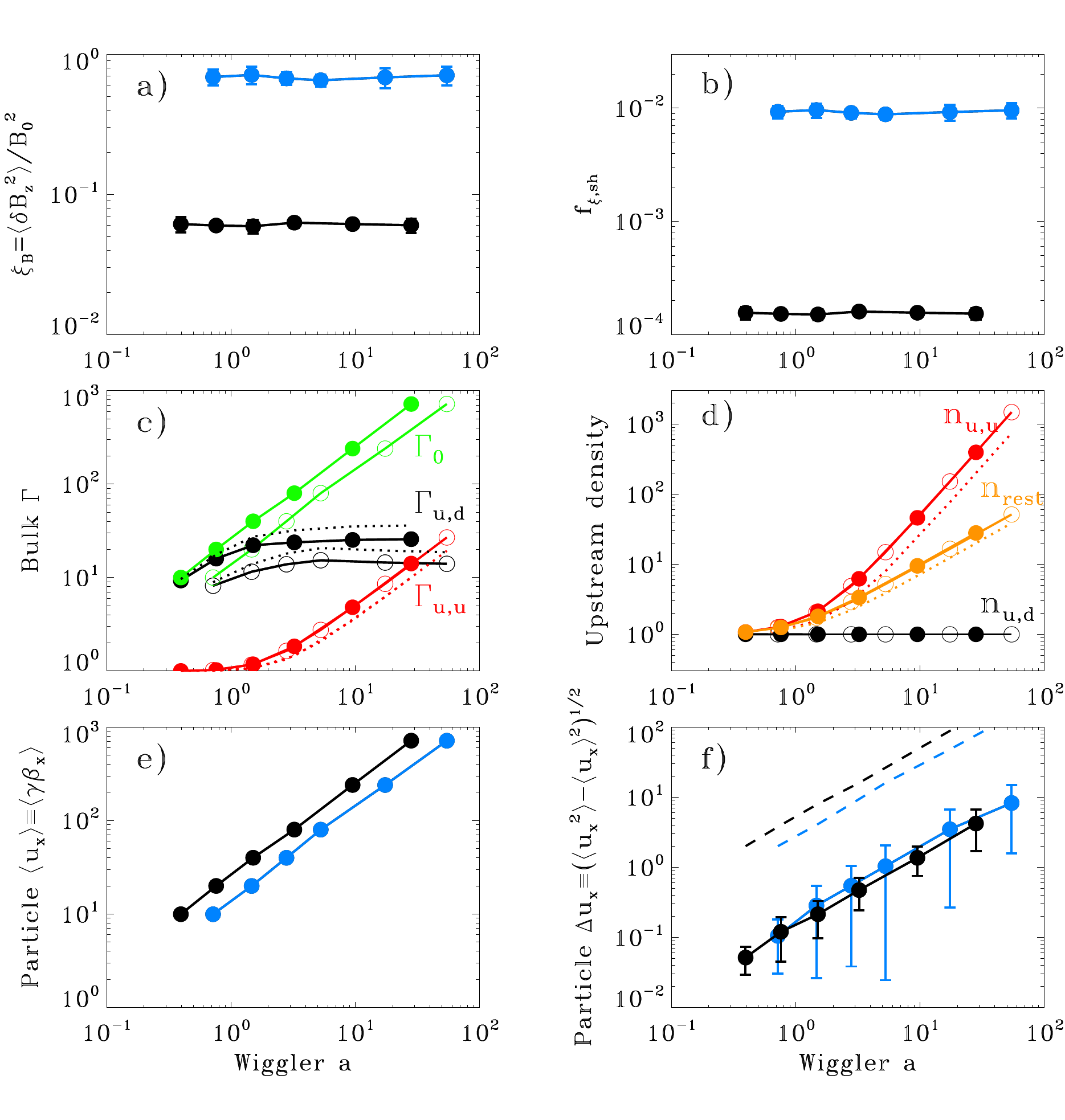}
    \caption{Results from a suite of 1D PIC simulations of magnetized shocks propagating into an upstream cold ($kT/m c^2=10^{-6}$) pair plasma of (initial) bulk Lorentz factor $\Gamma_0$ (the simulation is performed in the post-shock frame) and magnetization $\sigma = 3$ ({\it black points}) and $\sigma = 0.3$ ({\it blue points}) as a function of the wave-strength (wiggler) parameter $a$ (eq.~\ref{eq:strength}; in the simulation setup we vary the upstream Lorentz factor $\Gamma_0$, which results in changes in $a \propto \Gamma_0$, see green lines in panel c) of the electromagnetic precursor, i.e. the nascent FRB.  Quantities shown include: (a) fractional magnetic energy density of the precursor wave relative to the incident magnetic energy density of the upstream medium; (b) efficiency of the precursor electromagnetic wave, as measured in the shock frame; (c) various bulk Lorentz factors, where in this panel the filled circles correspond to $\sigma = 3$ and open circles to $\sigma = 0.3$.  In addition to the initial Lorentz factor of the far upstream medium ($\Gamma_0$), we show the actual Lorentz factor of the upstream medium just ahead of the shock resulting from its deceleration by the precursor wave, as measured in the downstream ($\Gamma_{\rm u, d}$) and initial upstream frame ($\Gamma_{\rm u,u}$); 
    (d) particle density in the region decelerated by the precursor wave, normalized by the initial (unperturbed) upstream density. 
    We find that, as measured in the downstream (simulation) frame, $n_{\rm u,d}$ is unaffected by the strong precursor wave (black circles; open, filled for $\sigma = 0.3$, $3$, respectively). This implies an increased density (``compression'') as measured in the rest frame of the decelerated electrons ($n_{\rm rest}$; orange) and in the unperturbed far-upstream frame ($n_{\rm u,u}$; red); 
    (e) mean longitudinal particle momentum in the precursor region, which remains $\sim \Gamma_0$ regardless of $a$; (f) dispersion in longitudinal particle momentum in the precursor region.  Dotted lines in panels c and d show for comparison the theoretical expectations for the effect of a strong-wave on the upstream medium from \citet{Lyubarsky19}. In the last panel, dashed lines represent the critical dispersion where thermal effects are expected to suppress the precursor efficiency (Babul \& Sironi, in prep.). All the data points are obtained by averaging first in a spatial region between 5 and 55 plasma skin depths ($=c/\omega_{\rm p}$, where $\omega_{\rm p}$ is the plasma oscillation frequency) ahead of the shock (i.e., across multiple precursor wavelengths) and then over time from $2000\,\omega_{\rm p}^{-1}$ and $6000\,\omega_{\rm p}^{-1}$.}
    \label{fig:PIC}
\end{figure*}

We also find that the radiative efficiency of the precursor emission is not changed for $a \gg 1$ from the value in the $a \ll 1$ limit, at least for the range $a \lesssim 60$ we have thus far explored (Fig.~\ref{fig:PIC}, panels a,b).  
This result is somewhat counter-intuitive, as one might expect deceleration of the upstream due to interaction with the strong wave (Fig.~\ref{fig:PIC}, panel c) to reduce the kinetic shock power and resulting maser efficiency. However, as discussed in Appendix~\ref{sec:appendix_strengthparameter}, we find this is balanced by compression of the upstream medium (Fig.~\ref{fig:PIC}, panel d) and transverse ``heating'' of the upstream electrons.
This heating could itself
stifle the maser by preventing the development of the necessary ring-shaped momentum distribution of particles immediately behind the shock.  However, as shown in the bottom panels of Fig.~\ref{fig:PIC}, the majority of the particle heating occurs in the plane transverse to the direction of wave propagation (parallel to the shock front) and therefore does not act to suppress the maser (see Appendix~\ref{sec:appendix_strengthparameter} for additional discussion).

Some degree of ``longitudinal'' heating of the upstream particles of the kind detrimental to the maser does occur, and is seen to increase with $a$  (Fig.~\ref{fig:PIC}, panel f).\footnote{We remark that at a given distance ahead of the shock, in 1D simulations the particle distribution has negligible dispersion. The spread in parallel momentum shown in Fig.~\ref{fig:PIC} comes from averaging over several precursor wavelengths.}  Extrapolating our results to yet higher values of $a \approx \Gamma $, the longitudinal heating is predicted to approach the level that Babul \& Sironi (in prep) find would begin to suppress the maser efficiency.  All else being equal, the strength-parameter of the wave scales with the maser efficiency as $a \propto f_{\xi}^{1/2}$ (eq.~\ref{eq:Appendix_a0}).  We speculate that for $a \gg \Gamma$ the maser emission will ``self-regulate'', resulting in the efficiency being capped at a maximum value $f_{\xi, \rm max} \sim 10^{-3}(f_{\rm e}/0.5)(m_*/m_{\rm e})^{-1}$ (see eq.~\ref{eq:fximax} in Appendix~\ref{sec:appendix_strengthparameter}).  

\subsection{Inferences about the FRB Engines}
\label{sec:engine}

Our results in \S\ref{sec:model} as presented in Figs.~\ref{fig:properties}-\ref{fig:GRB} are relatively independent of the nature of the central engine giving rise to the relativistic flares.  Our findings are also largely insensitive to the precise composition of the upstream medium as quantified by the values of $f_{\rm e}$ and $m_*$.  In this section we apply these findings to draw inferences about the population of FRB engines, before returning to the nature of the upstream medium in $\S\ref{sec:upstream}$.

The absolute energy loss rates implied by Fig.~\ref{fig:Edot} of $\dot{E} \sim 10^{39}-10^{41}$ erg s$^{-1}$ are significant.  For a source age of $\gtrsim 8$ years (minimum active age of the first repeater FRB~121102), the implied total engine energy budget is $E_{\rm tot} \gtrsim 10^{47}-10^{49}$ erg.  What engines could produce such values of $E_{\rm tot}$ and $\dot{E}$, a large part of which is placed into irregular ultra-relativistic outbursts?  

The inferred values of $\dot{E}$ are, for instance, similar to the Eddington luminosity $L_{\rm edd} \sim 10^{41}(M/10^{3}M_{\odot})$ erg s$^{-1}$ for a $M\sim 10^{2}-10^{4}M_{\odot}$ compact object.  Ultra-luminous X-ray sources (ULX) are postulated to arise in part from massive stellar-mass, or intermediate-mass, neutron stars or black holes which are accreting near or above the Eddington luminosity \citep{Kaaret+17}.  ULXs appear to occur across a range of galaxy Hubble types \citep{Swartz+11}, consistent with the current sample of FRB hosts.  Compact objects can produce highly relativistic jetted outflows containing a significant fraction of the accretion power (e.g.~\citealt{Mirabel&Rodriguez99}).  On the other hand, these transient's jets rarely obtain $\Gamma \gg 100$, except potentially in the more extreme case of cosmological gamma-ray bursts.  The nature of the time-dependent upstream medium into which the flare ejecta propagates is also not obvious in a ULX scenario.

The most actively discussed FRB engine model is a flaring magnetar (e.g.~\citealt{Popov&Postnov13,Lyubarsky14,Kulkarni+14}) at a particularly hyper-active phase early in its life (\citealt{Beloborodov17,Metzger+17}).  The magnetar could be formed in a core collapse supernova, through the merger of a binary neutron star system, or via the accretion-induced collapse (AIC) of a white dwarf \citep{Margalit+19,Beloborodov19}.  The required energy budgets are consistent with the magnetic energy reservoir of a magnetar, $E_{B_{\star}} \sim 3\times 10^{49}B_{16}^{2}$ erg, where $B = B_{16} \times 10^{16}$ G is the initial internal magnetic field strength.  
The energy loss-rates are also consistent with the predicted power output due to the ambipolar diffusion of Poynting flux from the core of a young magnetar, $\dot{E}_{\rm mag} \sim 10^{39}-10^{41}B_{16}^{3}$ erg s$^{-1}$ (eq.~2 of \citealt{Margalit+19}) for the same value of $B_{16}$ \citep{Beloborodov&Li16,Beloborodov19}.\footnote{The prefactor in this expression depends on whether the core of a neutron star is cooled by the modified or direct Urca process, respectively \citep{Beloborodov&Li16}, which in turn depends on the mass of the magnetar.}  

When possible to measure, other FRBs have shown dimmer persistent source luminosity and lower RM values than FRB~121102.  The magnetar model predicts a range of persistent source fluxes and RM values, which depends on the age of the magnetar-birthing supernova or merger/AIC event (\citealt{Margalit+19}).  For an assumed power-law decay of the time-average power of the baryon wind into the nebula  $\dot{E}_{\rm w} \propto t^{-\alpha}$, \citet{Margalit&Metzger18} predict secular decay of the rotation measure ${\rm RM} \propto t^{-(6+\alpha)/2}$ and persistent source flux $F_{\nu} \propto t^{-(\alpha^{2}+7\alpha-2)/4}$
for $\alpha >1$.\footnote{In their Appendix, \cite{Margalit+19} also investigate the $\alpha=0$ case and find $F_\nu \propto t^{-143/160} \simeq t^{-0.89}$ in this regime.}
For example, even in the case of 
$\alpha \approx 1$, the nebular RM of a source 10 times the age of FRB 121102 would be $\approx 3000$ times lower and the persistent source $\approx 30$ times dimmer.

\subsection{Nature of the Upstream Medium}
\label{sec:upstream}

One of the key open questions in the magnetar external shock scenario is whether the upstream medium into which the flare ejecta collides is (1) direct material from the electron-ion shell (e.g.~from a previous magnetic flare); (2) an $e^{-}/e^{+}$ dominated medium generated by a rotationally-powered component of the wind operating between the FRB-producing magnetic flares.  The latter could still be confined into a nebula between the wind termination shock and the external ion shell (\citealt{Beloborodov19}).     

One relevant fact is that the intrinsic polarization angle of bursts from FRB~121102 was measured to be roughly constant over $\gtrsim 7$ months of observations \citep{Michilli+18}, during which the source presumably underwent thousands of bursts.  In the synchrotron maser scenario, this requires a fixed direction for the magnetic field of the upstream plasma into which the FRB-producing flare collides.  Such a fixed field structure is expected in the outflow from a rotating compact object, for which the magnetic field wraps around the approximately-fixed rotation axis (\citetalias{Metzger+19}, \citealt{Beloborodov19}).  Though the clear prediction of a rotationally-powered pulsar-like wind, even the magnetic field in an ion shell would be wrapped in the toroidal direction provided that the timescale over which its mass was released from the magnetar surface (e.g.~$T_{\rm ion} \sim 1 \, {\rm s}$, a value motivated by the observed durations of giant Galactic magnetar flares; e.g.~\citealt{Palmer+05}) is comparable to, or greater than, the spin period of a decades old magnetar.  The fixed polarization angle of the bursts alone does not therefore distinguish the upstream medium in the magnetar scenario.

Another question is whether a pulsar-like wind from the magnetar can inject sufficient pairs between its magnetic flares to generate a nebula of sufficient pair density to match those required by our modeling (Fig.~\ref{fig:properties}).  In Appendix~\ref{sec:appendix_PWN} and Figure~\ref{fig:Ne} we show that the answer to this question depends on the assumed pair loading of the wind (e.g. whether the pair loading is some multiple of the standard \citealt{Goldreich&Julian69} particle flux or the different prescription advocated by \citealt{Beloborodov19}). However, for fiducial parameters --- the rotationally-powered wind cannot provide the upstream electron density inferred from observations.

A separate and potentially serious problem in invoking the relativistic, rotationally-powered pulsar wind as the source of the upstream medium is that the pairs, once shock heated at the wind termination shock, would be too hot to generate efficient maser synchrotron emission (\citealt{Beloborodov19}).  Appendix \ref{sec:appendix_PWN} provides estimates for the cooling of such pairs by synchrotron radiation in the nebula and inverse Compton emission off the background radiation field provided by the incoherent synchrotron emission of the FRB-generating shocks.  Although synchrotron emission is found to be insufficient to cool the pairs between flares, inverse Compton cooling from the most energetic flares, $E_{\rm flare} \gtrsim 2 \times 10^{44}$ erg, may be sufficient to cool all the pairs in the nebula to sub-relativistic temperatures.  However, if such cooling events are required to occur too frequently (e.g. as the shortest observed burst wait-times $\sim 10^{2}$ s), the energy budget of the magnetar would be strained.  On the other hand, one cannot completely exclude that rare powerful flares create a sufficiently large ``reservoir'' of cold electrons to enable efficient maser emission long after.  In summary, although invoking a magnetar wind pair nebula as the upstream medium holds advantages relative to the ion-loaded shell, it also comes with several issues and uncertainties.

\subsection{Comparison to Beloborodov (2019)}
\label{sec:Beloborodov}

Although the picture presented in \citetalias{Metzger+19} and here are broadly similar to that of \citet{Beloborodov19}, a number of key differences separate our models.  While we assume a slowly-expanding, relatively mildly magnetized medium with a potentially substantial ion component, \citet{Beloborodov19} invokes the upstream medium being a high $\sigma \gg 1$ ultra-relativistic, pair-loaded spin-down powered wind from the magnetar.  As one consequence of this, the shock radii in the \citet{Beloborodov19} scenario are larger $\sim 10^{14}$ cm than in our picture, which are instead in the range $r_{\rm sh} \sim 10^{12}-10^{13}$ cm at the time of the observed bursts.  For similar reasons, while induced Compton scattering plays an important role in shaping the observed FRB light curve and SED in our scenario (particularly generating a low frequency cut-off), it plays little or no role in the \citet{Beloborodov19} scenario.  


The repeating FRB sample show a range of frequency decay rates (Fig.~\ref{fig:driftrate}), which we have shown in the \citetalias{Metzger+19} scenario can be matched to reasonable diversity in the power-law slope $k$ of the upstream medium (e.g.~tail end of the ion shell; Fig.~\ref{fig:cartoon}) and would naturally be expected to evolve between bursts.  Given that the density slope of a steady pulsar wind would naively be identical in every case, it is not clear to us how similar diversity of the drifting rate would be imprinted if the upstream medium is a relativistic pulsar wind.  We also note that matching the frequencies and fluences of observed FRBs within the \cite{Beloborodov19} model requires a high particle flux being ejected shortly preceding flares.  We find that this is only possible if the pulsar wind preceding the flares is highly mass-loaded and therefore only mildly relativistic and moderately magnetized (Lorentz factor $\Gamma_{\rm w} \sim 4$, magnetization $\sigma \sim 2$), in which case the pulsar wind is not terribly different in its properties than the effectively stationary medium assumed here and by \citetalias{Metzger+19}.


\section{Conclusions}

We have modeled the sample of repeating FRBs, including the newly discovered population by CHIME \citep{CHIME_newrepeaters}, within the synchrotron maser decelerating blast wave model of \citetalias{Metzger+19}, which in turn built on pioneering work by \citet{Lyubarsky14} and  \citet{Beloborodov17}.  Our main results as summarized in Figs.~\ref{fig:driftrate}-\ref{fig:properties} are measurements, for each burst, of the energy $E_{\rm flare}$ and minimum Lorentz factor $\Gamma$ of the flare ejecta; the radius $r_{\rm sh}$ of the shock interaction at the time of the observed FRB; and the value $n_{\rm ext}$ and radial power-law index $k$ of the density of the external medium at $r_{\rm sh}$.  From the inferred burst energetics and rates, we also place loose constraints on the required power output of the central engine, $\dot{E}$ (Fig.~\ref{fig:Edot}).

Several consistency checks are applied which lend credence to the model.  The values of $k \in [-2,1]$ we find from the observed downward drift rate of FRB sub-pulses are consistent with our assumption of blastwave deceleration using the Blandford-McKee self-similar dynamics.  Furthermore, the dynamical timescales at the shocks are inferred to be short compared to the wait time between the vast majority of the flares (Fig.~\ref{fig:waittime}), thereby justifying a sufficiently cold medium by the time of the next shock for efficient maser emission.  Our inference of shock radii $\sim 10^{12}-10^{13}$ cm are also consistent with the hypothesis that the upstream medium is a dense baryon-loaded shell from the most powerful flares, which take place every $\sim 10^{4}-10^{5}$ s when the source is active (Fig.~\ref{fig:cartoon}).  Indeed, the power output in such baryonic ejecta appears broadly similar to that of the ultra-relativistic FRB-producing ejecta (Fig.~\ref{fig:Edot}), consistent with a common origin.  

For individual bursts other than FRB 121102, our results depend on the uncertain source distance, for which only rough upper limits can be placed from the source DM.  Our results are relatively insensitive to uncertain parameters of the model, such as the composition of the upstream medium (particularly the value of $m_*$ setting the peak frequency of the maser emission).  An important exception is the maser efficiency $f_{\xi}$ to which the flare energy is roughly inversely proportional (eq.~\ref{eq:Eflare}).  The value of $f_{\xi}$ is reasonably well constrained theoretically for an $e^{-}/e^{+}$ plasma (e.g.~\citealt{Plotnikov&Sironi19}).  

One possible concern about the maser mechanism is the large predicted strength/wiggler parameter $a \gg 1$ of the FRB wave
and its impact on the immediate-upstream
(Fig.~\ref{fig:strengthparameter}).  We have presented 1D PIC simulations of magnetized shocks which show (to our knowledge for the first time) that the efficiency of the shock maser emission is not appreciably suppressed in the case of $e^{-}/e^{+}$ upstream composition even in cases with $a \gg 1$ (Fig.~\ref{fig:PIC}).  We confirm the theoretical expectation that the upstream electrons are accelerated to relativistic velocities by the electric field of the wave (both in their bulk gyrocenter motion, and internally).  However, most of the electron ``heating'' occurs transverse to the direction of the wave propagation and therefore does not destroy the post-shock ring-shaped momentum distribution of the electrons which is critical for the maser emission. 

The simulations do show evidence for appreciable longitudial heating by the wave, which when extrapolated to yet stronger waves than those we have simulated could act to suppress the maser efficiency for $a \gtrsim \Gamma$.  In such a regime, we speculate that the upstream heating will act to self-regulate the maser, capping the FRB radiative efficiency of a blastwave undergoing self-similar deceleration at a maximum value of $f_{\xi,\rm max} \approx 10^{-3}(f_{\rm e}/0.5)(m_*/m_{\rm e})^{-1}$ (eq.~\ref{eq:fximax}; Appendix \ref{sec:appendix_strengthparameter}).   Future simulation work will more thoroughly explore the properties of the precursor wave generation in the $a \gg \Gamma$ limit, both for the $e^{-}/e^{+}$ and electron-ion cases. 

For purposes of economy, we have nominally favored an electron-ion composition for the upstream medium surrounding the FRB engine based on the requirement to reproduce the persistent synchrotron emission and high RM of FRB~121102 (provided both arise from an engine-fed nebula on larger scales; e.g.~\citealt{Margalit&Metzger18}).  However, large uncertainties remain regarding the efficiency and SED of the maser emission in the electron-ion case (most PIC simulation work has focused on the $e^{-}/e^{+}$ case due to computational limitations) and we cannot exclude that the ion shell from a magnetar flare would possess an appreciable $e^{-}/e^{+}$ component.  An $e^{-}/e^{+}$ upstream medium could also be created by the rotatationally-powered wind of a pulsar or magnetar engine between flares (see also \citealt{Beloborodov19}).  A major uncertainty in this case is whether the pairs in the nebula have sufficient time to cool between flares to allow the maser to operate (Appendix \ref{sec:appendix_PWN}).  It is also unclear whether the pair multiplicity of the pulsar wind is sufficiently high to explain the number of pairs required to explain the FRB data (Fig.~\ref{fig:Ne}).

One implication of our prediction that the peak of the FRB spectrum drifts downward in frequency due to blast wave deceleration is that multi-band  detections of a single burst may be possible, e.g. with parallel observing campaigns.  As a result of the increasing blast wave radius $r$, and decreasing Lorentz factor $\Gamma$ with time, the observed burst duration $\sim r/\Gamma^2$ will be intrinsically longer at lower frequencies (i.e. once the effects of scattering broadening$-$which is greater at low frequencies$-$has been removed).    

Some FRBs with time-resolved spectra show no apparent downward frequency drift.  Such zero-drifting behavior ($\beta \approx 0$) is possible in our model under some circumstances, e.g. $k \approx -2/7$ in the $t \lesssim \delta t$ case (Fig.~\ref{fig:driftrate}).  Indeed, for radial density profiles $k \lesssim -2/7$ and short bursts obeying $t \lesssim \delta t$, rising frequency behavior (``chirping'') is also possible.  Such a very short-lived chirp could in principle precede a longer sub-burst generated at times $t \gtrsim \delta t$ by the SED re-entering the observational band-pass from above and which would exhibit the standard downward-drifting behavior.  

Our results for the burst timescale, energetics, and rates are consistent with the hypothesis that repeating FRBs arise from young, hyper-active flaring magnetars.  However, we emphasize that the methodology as presented is generally applicable to any central engines involving an impulsive injection of energy into a moderately dense, strongly magnetized external medium.  This flexibility in the modeling would become particularly relevant if additional evidence supports the hypothesis of distinct repeating and non-repeating classes of bursts or if magnetar engines become disfavored for a sub-fraction of the bursts, e.g. based on host galaxy demographics.

Nevertheless, we feel that it is premature to abandon FRB unification scenarios entirely.  It is true that, when possible to measure, other FRBs have shown dimmer persistent source luminosity than FRB~121102; for example, no persistent radio source is detected from the location of recently-localized ASKAP FRB 180924 to a limit $\sim 3$ times lower than FRB~121102 and the bursts possess a much lower RM $\simeq$~14~rad~m$^{-2}$ (e.g.~\citealt{Bannister+19}).  However, as discussed in $\S\ref{sec:engine}$, such diversity could easily be attributed to (1) a lower rate of mass injection into the nebula (e.g. due to less energetic flaring activity) or (2) a different medium confining the nebula.  An example of the latter would be magnetars formed from binary neutron star mergers, for which the (lower mass) ejecta shells would expand more rapidly than in a supernova (e.g.~\citealt{Waxman17,Margalit+19}).  

One property previously argued to separate the repeating FRB population from the apparent one-off population is the longer observed duration of the repeating FRBs \citep{Scholz+16,CHIME_newrepeaters}.  We speculate that this is related to a higher rate of flaring by the repeating sources, which makes them more likely to be identified as repeaters and might arise physically from intrinsic diversity in the engine properties (e.g.~magnetars with strong internal magnetic fields and correspondingly shorter core ambipolar diffusion timescales for magnetic flux escape).  Inverting equation (\ref{eq:next}), the burst duration can be related to the density of the external medium according to $t \propto n_{\rm ext}^{3/2}$.  If each major flare were to ejecta a similar baryonic mass, then we would expect $n_{\rm ext}$ and thus the observed burst duration to decrease with the engine flaring rate.  

\section*{Acknowledgements}

We thank Andrei Beloborodov for helpful comments and conversations.  We thank Yuri Lyubarsky for sharing an early version of his manuscript on induced Compton scattering.
We thank Ryan Shannon, Xavier Prochaska, and Keith Bannister for providing information regarding FRB~181112.
BM thanks the Aspen Center for Physics where part of this work was completed. The Aspen Center for Physics is supported by National Science Foundation grant PHY-1607611. 
BM is supported by the U.S. National Aeronautics and Space Administration (NASA) through the NASA Hubble Fellowship grant \#HST-HF2-51412.001-A awarded by the Space Telescope Science Institute, which is operated by the Association of Universities for Research in Astronomy, Inc., for NASA, under contract NAS5-26555.
BDM acknowledge support from the Simons Foundation through the Simons Fellows Program (grant number 606260). LS and BDM acknowledge support from NASA ATP 80NSSC18K1104.




\appendix
\section{Derivation of Flare Properties}
\label{sec:appendix}

Here we briefly derive expressions for the flare energy, Lorentz factor, and upstream density in the \citetalias{Metzger+19} model as a function of the observed burst properties.
The peak frequency of the synchrotron maser SED is \citep{Plotnikov&Sironi19}
\begin{equation} \label{eq:Appendix_nupk}
    \nu_{\rm pk} \sim 3 \Gamma \nu_{\rm p} \propto m_*^{-1/2} f_{\rm e}^{1/2} n_{\rm ext}^{1/2} \Gamma.
\end{equation}
In the above equation $\nu_{\rm p}$ is the plasma frequency of species with effective mass $m_*$ and $f_{\rm e}$ is the ratio of electron to ion density.
Induced Compton scattering of the burst in the upstream medium scatters low-frequency photons out of the radiation cone so that the observed spectrum peaks at a frequency higher than $\nu_{\rm pk}$, where the induced Compton optical depth drops below $\tau_{\rm ICS} \sim 3$.
This implies (eq.~43 of \citetalias{Metzger+19})
\begin{equation} \label{eq:Appendix_taumax}
    3 \sim \tau_{\rm ICS} \left( \nu_{\rm max} \right) \propto \frac{f_{\rm e} n_{\rm ext} t}{r^2} \left. \nu^{-2} L_\nu \right\vert_{\nu_{\rm max}} 
    \propto f_{\rm e} \Gamma^4 n_{\rm ext}^2 \nu_{\rm pk}^{-3} \left( \frac{\nu_{\rm max}}{\nu_{\rm pk}} \right)^{-\alpha} t
    ,
\end{equation}
where in the last line we have 
assumed an intrinsic synchrotron maser SED of $\nu L_\nu \propto \nu^{3-\alpha}$ for $\nu \gtrsim \nu_{\rm pk}$ (our fiducial value throughout the paper is $\alpha \approx 4$, consistent with moderate $\sigma$; \citealt{Plotnikov&Sironi19}),
and used the fact that (eq.~12 of \citetalias{Metzger+19})
\begin{equation}
    \left. \nu L_\nu \right\vert_{\nu_{\rm pk}} \approx f_\xi L_{\rm sh} = f_\xi 4\pi r^2 n_{\rm ext} \Gamma^4 m_{\rm p} c^3 .
\end{equation}
Using the above expression we write the observed burst energy as
\begin{equation} \label{eq:Appendix_epsilon}
    \varepsilon \approx t \left. \nu L_\nu \right\vert_{\nu_{\rm pk}} \left(\frac{\nu_{\rm max}}{\nu_{\rm pk}}\right)^{3-\alpha}
    \propto \Gamma^8 n_{\rm ext} t^3 \left(\frac{\nu_{\rm max}}{\nu_{\rm pk}}\right)^{3-\alpha}
    ,
\end{equation}
where in the final line we have used $r \approx 2 \Gamma^2 c t$.

Equations~(\ref{eq:Appendix_nupk},\ref{eq:Appendix_taumax},\ref{eq:Appendix_epsilon}) can be combined to solve for the flare Lorentz factor $\Gamma$ and upstream external medium $n_{\rm ext}$ at the time $t$ at which the FRB is detected, and as a function of the observed burst frequency $\nu_{\rm max}$, duration ($\delta t$ and/or $t$ if $t \gtrsim \delta t$), and fluence (which enters $\varepsilon$; eq.~\ref{eq:epsilon}).
This results in the following expressions for the Lorentz factor,
\begin{align}
    \Gamma &= \left( \frac{16 \pi^2 m_{\rm e} m_{\rm p} c^5}{9 e^2}\right)^{-1/6} \left( \frac{6480 e^4}{\pi \sigma_{\rm T} m_{\rm e} m_{\rm p} c^4} \right)^{-\frac{\alpha-1}{6(\alpha+1)}} 
    \\ \nonumber
    &\times
    \left( \frac{m_*}{m_{\rm e}} \right)^{\frac{\alpha-3}{6(\alpha+1)}}
    f_{\rm e}^{\frac{1}{3(\alpha+1)}} f_\xi^{-\frac{1}{3(\alpha+1)}} \nu^{-\frac{\alpha+3}{6(\alpha+1)}} \delta t^{-\frac{1}{3(\alpha+1)}} t^{-1/3} \varepsilon^{1/6} 
    ,
\end{align}
and the external density
\begin{align}
    n_{\rm ext} &= \frac{\pi m_{\rm e}}{9 e^2} \left( \frac{16 \pi^2 m_{\rm e} m_{\rm p} c^5}{9 e^2}\right)^{1/3} \left( \frac{6480 e^4}{\pi \sigma_{\rm T} m_{\rm e} m_{\rm p} c^4} \right)^{\frac{\alpha+5}{3(\alpha+1)}} 
    \\ \nonumber
    &\times
    \left( \frac{m_*}{m_{\rm e}} \right)^{\frac{2\alpha-6}{3(\alpha+1)}}
    f_{\rm e}^{-\frac{3\alpha-1}{3(\alpha+1)}} f_\xi^{-\frac{4}{3(\alpha+1)}} \nu^{\frac{7\alpha+3}{3(\alpha+1)}} \delta t^{-\frac{4}{3(\alpha+1)}} t^{2/3} \varepsilon^{-1/3} .
\end{align}
In the above equations we have kept $\delta t$ separate from $t$, relevant in the regime where $t \lesssim \delta t$. In the opposite regime ($t \gtrsim \delta t$) the expressions should be modified substituting $\delta t \to t$. In this regime, $t$ should be associated with the observed burst duration (width).

Finally, the energy of the relativistic flare can be expressed as a function of the Lorentz factor and external density found above,
\begin{equation}
    E_{\rm flare} = 64\pi m_{\rm p} c^5 \Gamma^8 n_{\rm ext} t^3 
    \begin{cases}
    {\delta t}/{t} , &t \lesssim \delta t
    \\
    {2}/{(17-4k)} , &t \gtrsim \delta t
    \end{cases}{}
\end{equation}
which leads to the following solution
\begin{align}
    E_{\rm flare}(t \lesssim \delta t) = &\left( \frac{6480 e^4}{\pi \sigma_{\rm T} m_{\rm e} m_{\rm p} c^4} \right)^{-\frac{\alpha-3}{\alpha+1}} 
    \\ \nonumber
    &\times
    \left( \frac{m_*}{m_{\rm e}} \right)^{\frac{2(\alpha-3)}{\alpha+1}}
    f_{\rm e}^{-\frac{\alpha-3}{\alpha+1}} f_\xi^{-\frac{4}{\alpha+1}} \nu^{\frac{\alpha-3}{\alpha+1}} \delta t^{-\frac{4}{\alpha+1}} t^{0} \varepsilon .
\end{align}
The shock radius is easily obtained from the above equations, since $r_{\rm sh} \approx 2 \Gamma^2 c t$.

The inferred source properties as calculated above depend on the (for the most part unknown) redshift $z$ and luminosity distance $D$ of the source as
\begin{align} \label{eq:Appendix_zD_scaling}
&E_{\rm flare} \propto (1+z)^0 D^2
,~~
&n_{\rm ext} \propto (1+z)^{5/3} D^{-2/3},
\\ \nonumber
&\Gamma \propto (1+z)^{1/6} D^{1/3}
,
&r_{\rm sh} \propto (1+z)^{-2/3} D^{2/3} .
\end{align}
While the redshift effect is only minor for $z \lesssim 1$, the distance can have a significant impact. If the distance $D$ used to infer the source properties is an upper-limit on the true source distance (as is likely the case for our DM-inferred distances) then the inferred flare energy, Lorentz factor, and shock radius are only upper bounds on the true parameters, whereas the implied external density is a lower limit on the true density.

We conclude this section by deriving the frequency drift-rate $\beta$ (defined such that $\nu_{\rm max} \propto \nu^{-\beta}$; cf. eq.~\ref{eq:beta}) in the most general setting of the snychrotron blastwave model.
As discussed previously, \citetalias{Metzger+19} show that the observed peak frequency $\nu_{\rm max}$ is higher than the intrinsic peak maser frequency $\nu_{\rm pk}$ due to suppression by induced Compton scattering ($\tau_{\rm ICS}(\nu_{\rm pk}) \gg 1$).
Using the previously defined parameterization of the intrinsic maser SED, the induced Compton scattering optical depth at $\nu_{\rm max}$ can be expressed as $\tau_{\rm ICS}(\nu_{\rm max}) = \tau_{\rm ICS}(\nu_{\rm pk}) \times (\nu_{\rm max}/\nu_{\rm pk})^{-\alpha}$, and $\nu_{\rm max}$ is defined by the condition $\tau_{\rm ICS}(\nu_{\rm max}) \sim 3$ (see eq.~\ref{eq:Appendix_taumax}), so that
\begin{equation}
    \label{eq:Appendix_nupk}
    \nu_{\rm max}(t) \propto \nu_{\rm pk}(t) \times \tau_{\rm ICS}(\nu_{\rm pk})[t]^{1/\alpha} .
\end{equation}{}
The ICS optical depth at peak is related to the shock luminosity and upstream density as
\begin{equation}
\label{eq:Appendix_taupk}
    \tau_{\rm ICS}(\nu_{\rm pk}) \propto t \nu_{\rm pk}^{-3} L_{\rm sh} \frac{n_{\rm e}(r)}{r^2}
    \propto
    t \nu_{\rm pk}^{-3} L_{\rm sh} \times
    \begin{cases}
    r_{\rm sh}^{-(k+2)} &, k>-2
    \\
    const. &, k<-2
    \end{cases}
\end{equation}{}
where in the `standard' $k>-2$ case ICS is dominated by gas immediately ahead of the forward shock (and thus changes temporally as $r_{\rm sh}$ increases), while in the $k<-2$ case, i.e. for a steeply increasing density profile, the ICS optical depth is set by gas at large distances $\gg r_{\rm sh}$ and the $n_{\rm e}/r^2$ term is decoupled from the shock dynamics (and we therefore assume it is temporally constant over short timescales).

By combining eqs.~(\ref{eq:Appendix_nupk},\ref{eq:Appendix_taupk}) and using the temporal scaling of $\nu_{\rm pk}(t)$, $L_{\rm sh}(t)$, and $r_{\rm sh}(t)$ derived in \citetalias{Metzger+19} (their eqs.~37,38, 17,21, and 14,18, respectively), we find
\begin{equation}
\label{eq:Appendix_beta1}
    \beta(k>-2) = 
    \begin{cases}
    \frac{3k+\alpha k - 6 + 2\alpha}{2\alpha (4-k)}
    \\
    \frac{2k - 5 + 3\alpha}{2\alpha (4-k)}
    \end{cases}{}
    \underset{\alpha=4}{=}
    \begin{cases}
    \frac{7k+2}{8(4-k)} &, t \lesssim \delta t
    \\
    \frac{2k+7}{8(4-k)} &, t \gtrsim \delta t
    \end{cases}{}
\end{equation}{}
and
\begin{equation}
\label{eq:Appendix_beta2}
    \beta(k<-2) = 
    \begin{cases}
    \frac{-k+\alpha k - 14 + 2\alpha}{2\alpha (4-k)}
    \\
    \frac{3(\alpha-3)}{2\alpha (4-k)}
    \end{cases}{}
    \underset{\alpha=4}{=}
    \begin{cases}
    \frac{3(k-2)}{8(4-k)} &, t \lesssim \delta t
    \\
    \frac{3}{8(4-k)} &, t \gtrsim \delta t
    \end{cases}{}
    .
\end{equation}{}
These expressions can also be inverted to probe the density profile power-law index $k$ as a function of the (measureable) frequency drift-rate $\beta$ (e.g. Fig.~\ref{fig:driftrate}),
\begin{equation}
\label{eq:Appendix_k1}
    k( \beta ; t \lesssim \delta t) = 
    \begin{cases}
    \frac{8\alpha\beta+14+2\alpha}{2\alpha\beta+1-\alpha}
    \\
    \frac{8\alpha\beta+6-2\alpha}{2\alpha\beta+3+\alpha}
    \end{cases}
    \underset{\alpha=4}{=}
    \begin{cases}
    \frac{32\beta+22}{8\beta-3}
    &, \beta < -1/4
    \\
    \frac{32\beta-2}{8\beta+7}
    &, \beta > -1/4
    \end{cases}
\end{equation}
and
\begin{equation}
\label{eq:Appendix_k2}
    k( \beta ; t \gtrsim \delta t) = 
    \begin{cases}
    \frac{8\alpha\beta+9-3\alpha}{2\alpha\beta}
    \\
    \frac{8\alpha\beta+5-3\alpha}{2\alpha\beta+2}
    \end{cases}
    \underset{\alpha=4}{=}
    \begin{cases}
    \frac{32\beta-3}{8\beta}
    &, \beta < 1/16
    \\
    \frac{32\beta-7}{8\beta+2}
    &, \beta > 1/16
    \end{cases}
    .
\end{equation}

Note that eqs.~(\ref{eq:Appendix_beta1},\ref{eq:Appendix_beta2},\ref{eq:Appendix_k1},\ref{eq:Appendix_k2}) predict a temporally increasing frequency drift, $\beta<0$ (a ``chirp'' instead of a downward drift), if $k<-2/7 \simeq -0.29$ and $t \lesssim \delta t$.
This situation may be realized for especially energetic flares and may be preferentially observed at higher frequencies.\footnote{Note that eqs.~(\ref{eq:Appendix_beta1},\ref{eq:Appendix_beta2},\ref{eq:Appendix_k1},\ref{eq:Appendix_k2}) also predict a positive $\beta$ ``chirp'' if $k>4$, however the self-similar solutions used in deriving the above results break down for such large $k$ rendering this regime unphysical.}

\section{Strength Parameter \& Self-Regulation of Maser Emission}
\label{sec:appendix_strengthparameter}

The wave-strength (``wiggler'') parameter is defined as
\begin{equation}
    a = \frac{e E }{2\pi m_{\rm e}c \nu}
    \approx \left(\frac{e^2 L_\nu}{2 \pi^2 m_{\rm e}^2 c^3 r^2 \nu}\right)^{1/2},
\end{equation}
where $E$ and $\nu$ are the electric field and frequency of the FRB wave.  For $\nu = \nu_{\rm pk}$, at which the strength parameter is maximal, we find, using results from the previous section,
\begin{equation} \label{eq:Appendix_a0}
    a = \left(\frac{m_* f_\xi L_{\rm sh}}{18 \pi m_{\rm e}^2 c^3 f_{\rm e} \Gamma^2 r_{\rm sh}^2 n_{\rm ext}}\right)^{1/2} 
    \underset{{\rm BM}}{=} \left(\frac{2 m_{\rm p} m_* f_\xi}{9 m_{\rm e}^2 f_{\rm e}}\right)^{1/2} \Gamma
    ,
\end{equation}
where the final line makes use of the \cite{Blandford&McKee76} solution.  Figure \ref{fig:strengthparameter} shows that we infer $a \sim 100$ from the observed sample of FRBs, under our fiducial assumptions for the upstream medium ($m_* = m_{\rm e}$, $f_{\rm e} = 0.5$, $f_{\xi} = 10^{-3}$).  This result appears to raise questions about the validity of the \citetalias{Metzger+19} model, which was motivated by synchrotron maser emission properties derived from simulations in the $a \ll 1$ regime.  

A relativistically strong wave $a \gg 1$ can have several potential effects on the upstream medium.  First, the guiding center frame motion of upstream electrons/positrons will be accelerated to relativistic velocities $\Gamma_{\rm e} \sim a$ away from the shock (e.g.~\citealt{Arons72,Blandford72}).  The same mechanism is expected to ``heat'' the electrons in the guiding center rest frame to an internal Lorentz factor $\gamma_{\rm e} \sim a$ (\citealt{Lyubarsky19}).  Thus, even an initially cold upstream of electrons will find their energy boosted by a factor $\sim a^{2}$ as measured in the external stationary medium frame.  In light of these effects, one might expect the properties of the synchrotron maser to be drastically different in the $a \gg 1$ limit from the previously studied cases ($a \ll 1$).  However, we now discuss the results from 1D PIC simulations which suggest this is not the case for a moderately high strength parameter, at least for an $e^{-}/e^{+}$ upstream composition.

Following the numerical set-up of \citet{Plotnikov&Sironi19}, we have performed a suite of 1D PIC simulations of ultra-relativistic shocks propagating into an upstream pair plasma of magnetization $\sigma = 3$ and $\sigma = 0.3$, in order to explore at a preliminary level the effects of large strength parameter $a \gg 1$ on the dynamics of the upstream medium and the properties of the maser emission.  Figure \ref{fig:PIC} shows several quantities of interest to the shock dynamics and maser emission as a function of $a \propto \Gamma_0$, where $\Gamma_0$ is the initial Lorentz factor of the upstream medium (well ahead of the shock) in the downstream rest-frame.

Perhaps counterintuitively, our simulations in the regime $1 \ll a \ll \Gamma_0 $ do not show an appreciable drop in the efficiency of the maser emission (i.e. fraction of the far upstream kinetic+magnetic energy placed into precursor wave energy).  There are two reasons for this result, the first being kinematical.  In agreement with theoretical expectations (e.g.~\citealt{Lyubarsky19}) we find that the upstream electron fluid is indeed accelerated away from the shock at bulk Lorentz factor $\Gamma_{\rm e} \approx a$ in the original upstream rest-frame of the upstream ($\Gamma_{\rm u,u}$ in Fig.~\ref{fig:PIC}, panel c).  This reduces the Lorentz factor at which the shocked gas moves into the upstream from its original value $\Gamma$ to ($\Gamma_{\rm u,d}$ in Fig.~\ref{fig:PIC}, panel c)
\begin{equation}
    \tilde{\Gamma} \underset{\Gamma,\Gamma_{\rm e} \gg 1}{\approx} \frac{\Gamma^2 + \Gamma_{\rm e}^2}{2 \Gamma \Gamma_{\rm e}}
    \underset{\Gamma \gg \Gamma_{\rm e}}{\approx} \frac{\Gamma}{2 \Gamma_{\rm e}} \approx\frac{\Gamma}{2 a} .
\end{equation}
 The shock jump conditions dictate that the kinetic luminosity of the shock at fixed radius $r$ scales as $L_{\rm sh} \propto \Gamma^{2}n m c^{2}$, where $n$ is the rest-frame density of the upstream medium.  Our simulations show that the expected reduction in $\tilde{L}_{\rm sh} \propto \tilde{\Gamma}^{2} \propto 1/a^{2}$ from the bulk upstream acceleration is compensated by (1) an {\it increase} in the immediate upstream rest-frame density by a factor $\tilde{n}/n \approx a$ due to compression by the precursor wave ($n_{\rm rest}$ in Fig.~\ref{fig:PIC}, panel d); (2) the expected increase in the effective rest mass density due to the electrons being heated by the precursor wave to an internal $\tilde{m}/m \approx \gamma_{\rm e} \approx a$.  Thus, $\tilde{L}_{\rm sh} \approx L_{\rm sh}$, i.e. the shock luminosity (available electron energy) for otherwise similar far upstream parameters is similar to that in the $a \ll 1$ limit.  

A second non-trivial point relates to the effect of upstream heating on the efficiency of the precursor.  The maser emission is driven by the ring-shaped hole in the momentum space distribution of the post-shock electrons, which would not be created if the upstream plasma enter the shock too ``hot''.  More quantitatively, Babul \& Sironi (in prep) find that the maser efficiency appreciably drops from its value in the cold limit for upstream temperatures $kT \gtrsim kT_{\rm th} \approx 0.03 m_{\rm e} c^{2}$.  As expected, our simulations show that the precursor wave heats initially cold upstream electrons to an internal energy $\gamma_{\rm e} \sim a$, but the induced electron motion in the their gyrocenter frame is not isotropic (it cannot be thought of as a true ``temperature'').  In particular, the driven electron motion is mainly perpendicular to the direction of the wave propagation (parallel to the shock front) and thus will not suppress the maser.  

Instead, from our simulation results, we estimate that the spread in momentum parallel to the shock scales as $\delta u_{\parallel} \approx 0.15 a$ (Fig.~\ref{fig:PIC}; panel (f)).  Given the threshold $\delta u_{\parallel}/\Gamma_0 \gtrsim 0.2$ for efficient maser emission found by Babul \& Sironi (in prep), this implies that suppression of the maser emission will only take place for $a \gtrsim \Gamma$.  From equation (\ref{eq:Appendix_a0}) we see that for maser efficiencies above a critical threshold,
\be
f_{\xi} \gtrsim f_{\xi,\rm max} \sim 10^{-3}\times \left(\frac{f_{\rm e}}{0.5}\right)\left(\frac{m_*}{m_{\rm e}}\right)^{-1}
\label{eq:fximax}
\ee
we expect $a/\Gamma \gtrsim 1$ and therefore significant back-reaction of the precursor wave on the maser.  In such a limit one might expect a self-regulation of the maser process such that $f_{\xi} \approx f_{\xi,\rm max}$.  In other words,
\be
f_{\xi} \approx {\rm min}\left[f_{\xi,0}, f_{\xi, \rm max}\right] , 
\ee
where $f_{\xi, 0}$ is the maser efficiency in the $a \ll 1$ limit.

\section{Induced Compton Scattering}
\label{sec:appendix_ICS}

Induced Compton scattering (ICS) by electrons or positrons in the upstream medium ahead of the shock plays an important role in our model.  Though many of our results do not depend qualitatively on the details of this complex problem (or indeed on whether ICS is at all operational in our setting), the quantitative results may.
In the following we discuss a few subtleties and complications regarding ICS in our envisaged scenario. Although a full treatment of the ICS problem is outside the scope of this work, we show that the results derived in \citetalias{Metzger+19} are not greatly affected by several plausible complications.

ICS is an important effect for high brightness-temperature sources and has been investigated extensively in the context of pulsars and AGN (e.g.~\citealt{Melrose&Wilson72,Blandford&Scharlemann76,Sincell&Krolik92}). More recently \cite{Lyubarsky08} studied the role of ICS on short duration pulses relevant to FRBs. 
The effective optical depth that governs the degree that ICS down-scattering has on the emergent spectrum is roughly given by
\begin{eqnarray}
\label{eq:tauics}
    \tau_{\rm ICS} = \frac{3}{4} \tau_{\rm T} \frac{\partial}{\partial \nu} \left(\nu \frac{k T_{\rm b}(\nu)}{m_{\rm e} c^2}\right)
    \Omega_{\rm b}
    \times
    \begin{cases}
    c t / r &,\textrm{out-of-beam}
    \\
    \Omega_{\rm b} / 6 &,\text{in-beam}
    \end{cases}
    \\ \nonumber
    \approx \frac{3 \sigma_{\rm T} n_{\rm e} r}{8 \pi m_{\rm e}} \frac{\partial}{\partial \nu} \left(\frac{F_\nu}{\nu}\right) 
    \times 
    \begin{cases}
    1 / 2 \Gamma^2 &,\textrm{out-of-beam}
    \\
    \pi / 6 \Gamma^2 &,\text{in-beam}
    \end{cases}
    ,
\end{eqnarray}{}
where $\tau_{\rm T} = \sigma_{\rm T} n_{\rm e} r$ is the Thomson optical depth of the upstream medium of electron density $n_{\rm e}$ and radial extent $r$, and $T_{\rm b}$ is the brightness temperature of the incident radiation beam.  We have separated the scattering optical depths into an ``in-beam'' component which accounts for self-scattering of the collimated radiation, and ``out-of-beam'' scattering of the beam radiation into large angles $> \theta_{\rm b}$ (e.g. onto radiation seeded by spontaneous Thompson scattering out of the beam). The former is inhibited by the small beam solid angle $\Omega_{\rm b} \approx \pi \theta_{\rm b}^2$ because of the reduced electron recoil (and therefore the resulting scattering rate) for small angles, while the latter is dependent on the pulse duration $\sim c t$ rather than the physical extent of the scattering medium.
In its present application, the beam opening angle is $\theta_{\rm b} \sim 1/\Gamma$ due to the relativistic collimation, while the burst duration $\sim t$ is related to the emission radius via $r \approx 2 \Gamma^2 c t$, in which case we find that in-beam and out-of-beam scattering are of comparable strength up to a numerical factor of order unity.

The expressions used above are derived under the assumption of a cold (non-relativistic) plasma, but the ICS process depends on the electron distribution function \citep[e.g.][]{Wilson82}.  As pointed out by \cite{Beloborodov19}, the cold plasma assumption is not obviously applicable in our proposed scenario since the wave-strength parameter obeys $a \gg 1$ and thus electrons or positrons will be accelerated to relativistic velocities $\Gamma_{\rm e} \sim a$ away from the shock by the impinging wave (see Appendix~\ref{sec:appendix_strengthparameter}).

\citet{Lyubarsky19} treats ICS in the strong wave regime, finding that the scattering rate is suppressed by a factor $\sim 1/a^3$ in the $a \gg 1 $ limit relative to the standard $a \ll 1$ expression. This result was obtained for a stationary scattering medium, i.e. in the guiding center rest frame of the scattering electrons, and therefore must be transformed into the lab frame.  If such electrons acquire bulk motion with Lorentz factor $\Gamma_{\rm e}$ in the direction of the wave propagation, the appropriate induced Compton scattering optical depth $\tau_{\rm ICS}^\prime$ must be evaluated in the electrons' rest frame. Given the Lorentz transformation of the relevant quantities: $\tau_{\rm T}^\prime = \tau_{\rm T}$, $r^\prime/r = \nu^\prime/\nu = T_{\rm b}^\prime / T_{\rm b} = 1/\Gamma_{\rm e}$, $t^\prime = \Gamma_{\rm e} t$, $\Omega_{\rm b}^\prime = \Gamma_{\rm e}^2 \Omega_{\rm b}$ (e.g.~\citealt{Beloborodov19}) we find that $\tau_{\rm ICS}^\prime = \Gamma_{\rm e}^3 \tau_{\rm ICS}$.

Thus, if the ICS rate is indeed inhibited by $a^{-3}$ in the strong-wave regime ($a \gg 1$; \citealt{Lyubarsky19}) we finally obtain
\begin{equation}
    \tau_{\rm ICS} \xrightarrow{a, \Gamma_{\rm e} \gg 1} \tau_{\rm ICS} \left(\frac{\Gamma_{\rm e}}{a}\right)^3 .
\end{equation}{}
Thus, to the extent that electrons or positrons are accelerated away from the shock by the FRB to $\Gamma_{\rm e} \sim a$ (see Fig.~\ref{fig:PIC} and discussion in Appendix~\ref{sec:appendix_strengthparameter}), the effective ICS scattering rate ($\sim$ optical depth) would be unchanged with respect to the standard expression (which assumes $a \ll 1$ and $\Gamma_{\rm e} \approx 1$).

We conclude by clearing up a potential misconception about the applicability of ICS attenuation in our model.
The intrinsic SED of the synchrotron maser at frequencies above the peak $\nu_{\rm pk}$ approximately obeys $F_{\nu} \propto \nu^{-2}$ (e.g.~\citealt{Plotnikov&Sironi19}), in which case it would seem that ICS down-scattering would not appear to take place since $\tau_{\rm ICS} \propto \partial/\partial \nu(F_{\nu}/\nu) < 0$ at $\nu \gtrsim \nu_{\rm pk}$.  However, note that, just below the peak frequency the SED does obey $\partial/\partial \nu(F_{\nu}/\nu) < 0$ and therefore scattering will be highly effective since $\tau_{\rm C}(\nu_{\rm pk}) \gg 1$.  As low-frequency photons in the beam are attenuated by ICS moving out radially through the external medium, the value of $\nu_{\rm pk}$ will increase, and thus the frequency range over which $\partial/\partial \nu(F_{\nu}/\nu) < 0$ is satisfied and ICS can take place will increase.  This justifies our use of equation (\ref{eq:tauics}) 
with $\partial/\partial \nu \sim 1/\nu$
as a rough approximation to the true optical depth at all frequencies (uncertainty introduced by the intrinsic shape of the SED, which depends on the magnetization and temperature of the upstream medium, is likely to enter at a comparable level).  More work is needed to assess the detailed effects of ICS on the emergent photon spectrum.  

\section{Pulsar Wind as Upstream Medium}
\label{sec:appendix_PWN}

In what follows we consider whether material from a rotationally-powered wind from the magnetar, acting continuously in between the magnetic flares, could provide a suitable upstream environment for the FRB shocks (as an alternative to baryon loaded shells). The rotationally-powered wind would form a termination shock after colliding with the external baryon-loaded shell and the pairs would collect in a layer separating the magnetar from the inner edge of the shell.  Assuming an external homologously-expanding confining medium of density $n_{\rm sh}$ (see eq.~\ref{eq:mdot}), the radius of the nebula inflated by the wind of luminosity $L_{\rm sd}$ in the time $t_{\rm wait}$ between flares may be approximated as \citep[e.g.][]{Chevalier05}
\begin{align} \label{eq:Rneb}
R_{\rm n} &\sim \left(\frac{L_{\rm sd}t_{\rm wait}^{3}}{n_{\rm sh}m_{\rm p}}\right)^{1/5} \\ \nonumber
&\approx 2.7 \times 10^{12}\,{\rm cm} \, \left(\frac{t_{\rm age}}{30\,{\rm yr}}\right)^{-2/5}B_{\rm d,15}^{-2/5}\left(\frac{t_{\rm wait}}{100\,{\rm s}}\right)^{3/5}\left(\frac{n_{\rm sh}}{10^{4}\,{\rm cm^{-3}}}\right)^{-1/5},
\end{align}  
where in the second line we have substitute in the spin-down luminosity of the magnetar of age $t_{\rm age}$,
\be L_{\rm sd} \approx 2.5\times 10^{36}\left(\frac{t_{\rm age}}{30\,{\rm yr}}\right)^{-2}B_{\rm d,15}^{-2} \, {\rm erg \, s}^{-1}, \label{eq:Lsd} \ee
where $B_{\rm d} = B_{\rm d,15}\times 10^{15}\, {\rm G}$ and $t_{\rm age}$ are the surface dipole magnetic field and age of the magnetar, respectively.  We assume the magnetar age is much greater than the initial spin-down time $t_{\rm sd,0} = 114\,{\rm s}\,B_{\rm d,15}^{-2}P_{\rm 0,ms}^{-2}$, where $P_{\rm 0,ms}$ is the birth period in milliseconds.

\begin{figure}
    \centering
    \includegraphics[width=0.45\textwidth]{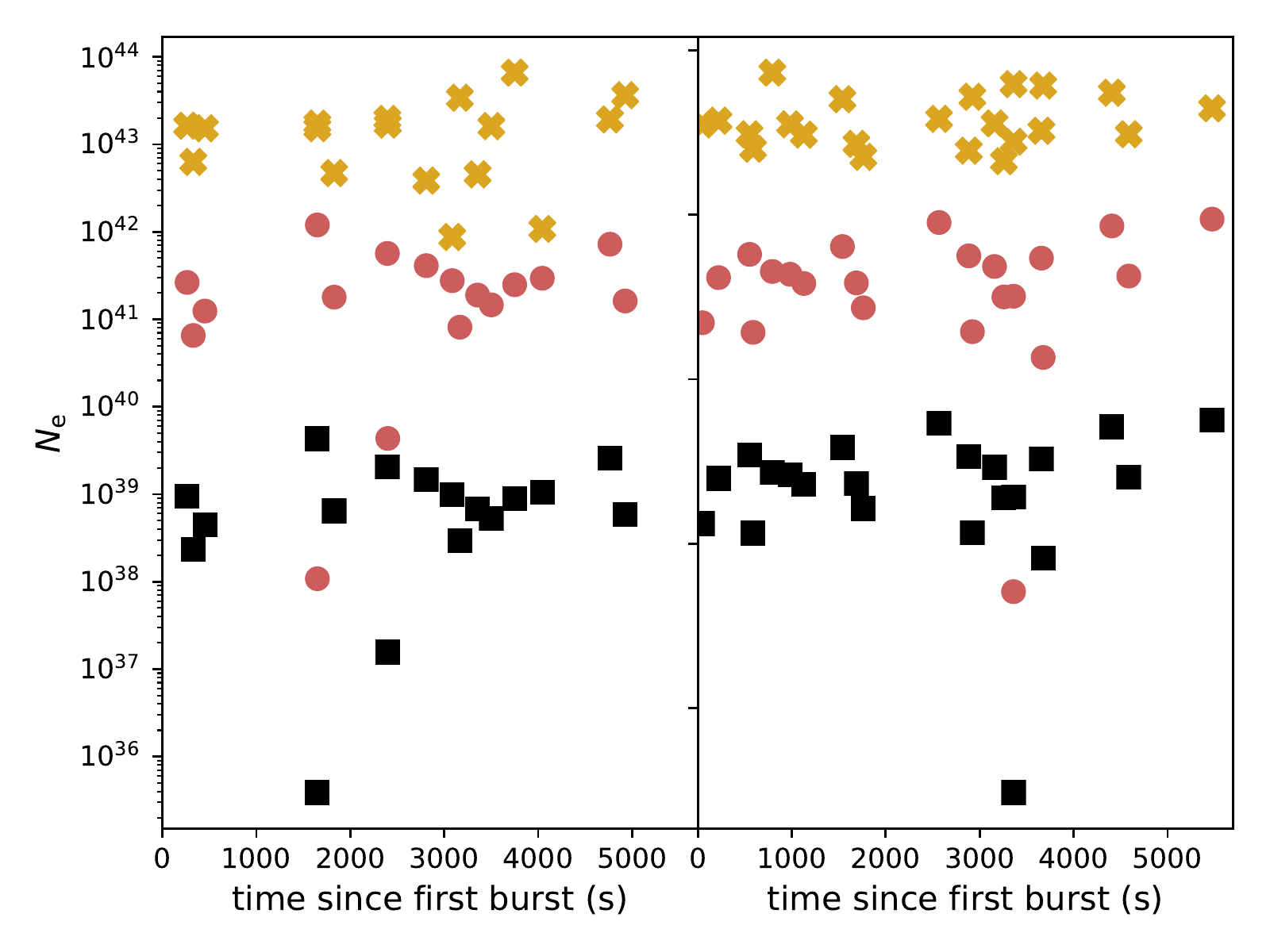}
    \caption{Number of required electrons $N_{\rm e}$ swept-up by the shocks producing the observed FRBs in the sample of \citet{Gourdji+19} assuming $f_{\rm e} = \mu_\pm = 10^3$ (yellow crosses; eq.~\ref{eq:AppendixNe}). This number should be compared with the number of pairs injected between successive bursts by a rotationally-powered wind (eq.~\ref{eq:Npm}). Black squares show the latter taking the pair loading scaled to the \citealt{Goldreich&Julian69} flux (eq.~\ref{eq:AppendixNdotGJ}), while red circles show the pair loading model of (\citealt{Beloborodov19}; eq.~\ref{eq:AppNdotBel}), each calculated for $\mu_{\pm} = 10^{3}$, $B_{\rm d} = 10^{15} \,{\rm G}$ and $t_{\rm age} = 30 \, {\rm yr}$. For fiducial parameters $N_\pm \ll N_{\rm e}$ indicating that the upstream medium cannot be supplied by a rotationally-powered wind.}
    \label{fig:Ne}
\end{figure}

The characteristic scale for particle loss in a rotationally powered wind is given by the \citet{Goldreich&Julian69} flux,
\begin{eqnarray} 
\dot{N}_{\pm} 
\equiv \mu_{\pm}\left(\frac{I}{e}\right) 
&\sim& 3\times 10^{43} \, {\rm s^{-1}}\left(\frac{\mu_{\pm}}{10^{3}}\right)B_{\rm d,15}P_{\rm ms}^{-2}, 
\nonumber \\
&\approx&  3.6\times 10^{36} \, {\rm s^{-1}}\left(\frac{\mu_{\pm}}{10^{3}}\right)B_{\rm d,15}^{-1}\left(\frac{t_{\rm age}}{{\rm 30\, yr}}\right)^{-1}
\label{eq:AppendixNdotGJ}
\end{eqnarray}
where $I \equiv \left. 4\pi c R_{\rm L}^{2}\eta_{\rm GJ}\right\vert_{\rm R_{\rm L}}$ and $\eta_{\rm GJ} \approx \Omega B/ 2\pi c$ is the Goldreich-Julian charge density evaluated at the light cylinder radius $R_{\rm L} = c/\Omega \simeq 48P_{\rm ms}\,{\rm km}$ and $\mu_{\pm}$ is the pair multiplicity of the wind.  
\citet{Beloborodov19} gives an alternative expression for the pair injection rate from a young magnetar (their eq.~12)
\begin{eqnarray} 
\dot{N}_{\pm}  
&\sim& 3\times 10^{42}\,{\rm s^{-1}} \, \left(\frac{\mu_{\pm}}{10^{3}}\right) P_{\rm ms}^{-1} B_{\rm d,15}^{2/3} 
\nonumber \\
&\approx& 1\times 10^{39} \, {\rm s^{-1}} \, \left(\frac{\mu_{\pm}}{10^{3}}\right) B_{\rm d,15}^{-1/3}\left(\frac{t_{\rm age}}{\rm 30 yr}\right)^{-1/2}.
\label{eq:AppNdotBel}
\end{eqnarray}
In the final lines of eqs.~(\ref{eq:AppendixNdotGJ}, \ref{eq:AppNdotBel}) we have substituted $P \simeq 2.88 \, {\rm s}\,B_{\rm d,15}(t_{\rm age}/{\rm 30 yr})^{1/2}$ for the magnetar spin period at times $t \gg t_{\rm sd,0}$ (e.g.~\citealt{Metzger+15}).  
The total number of pairs injected between baryon flares may therefore be estimated as
\begin{equation} 
N_{\pm} = \dot{N}_{\pm}t_{\rm wait}.
\label{eq:Npm}
\end{equation}
Figure~\ref{fig:Ne} compares $N_{\pm}$ to the number of electrons swept up by the shock, 
\begin{align}
\label{eq:AppendixNe}
    N_{\rm e} &= \frac{4\pi}{3} n_{\rm e} r_{\rm sh}^{3} 
    \\ \nonumber
    &\approx
    10^{41} \,
    \left(\frac{m_*}{m_{\rm e}}\right)^{1/3}
    \left(\frac{\min \left[ f_{\rm e}, \frac{m_{\rm p}}{m_{\rm e}} \right]}{0.5}\right)^{2/3}
    f_{\xi,-3}^{-2/3} \nuCHIME^{1/3} t_{-3} \varepsilon_{40}^{2/3}
    ,    
\end{align}{}
as implied by FRB~121102 bursts from the sample of \cite{Gourdji+19},
assuming an ``optimistic'' scenario where the upstream is pair-dominated by the magnetar wind, i.e. $f_{\rm e} = \mu_\pm = 10^3$ (yellow crosses).  
The number of pairs $N_\pm$ (eq.~\ref{eq:Npm}) that can be supplied by a Goldreich-Julian wind (eq.~\ref{eq:AppendixNdotGJ}) or a wind as prescribed by \cite[][eq.~\ref{eq:AppNdotBel}]{Beloborodov19} are shown as black squares and red circles, respectively.

Figure~\ref{fig:Ne} illustrates that the required upstream electrons cannot be fully provided by the rotationally-powered wind unless $(\mu_\pm/10^3) B_{\rm d,15}^{-1}(t_{\rm age}/30{\rm yr})^{-1} \gtrsim 10^4$ for a Goldreich-Julian wind, or
$(\mu_\pm/10^3) B_{\rm d,15}^{-1/3}(t_{\rm age}/30{\rm yr})^{-1/2} \gtrsim 10^2$ for the \cite{Beloborodov19} model.
If $\mu_\pm \sim 10^3$ is a reasonable value for the pair multiplicity then the latter would require an exceedingly weak magnetic field or young source age to be dominated by a rotationally-powered wind. This supports the hypothesis that the upstream medium in our scenario is set by baryon-loaded shells ejected between major flares.
Note, however, that we have here assumed that the relevant timescale for pair-pollution of the ion-shells is the wait time between adjacent bursts, $t_{\rm wait} \sim 10^2 \, {\rm s}$, however --- if the relevant timescale is instead determined by the time between major shell ejections, $\sim \Delta T \sim 10^5 \, {\rm s}$, then the the estimate of $N_\pm$ should be increased by a factor $\sim 10^3$ (eq.~\ref{eq:Npm}) in which case the \cite{Beloborodov19} model may indeed provide sufficient pairs.

As pointed out by \citet{Beloborodov19}, one concern in invoking material from the relativistic, rotationally-powered pulsar wind as the source of the upstream medium is that the pairs, once shock heated at the termination shock, would be too hot to generate synchrotron maser emission (Babul \& Sironi, in prep).  We check this assumption here by considering whether the shocked pairs will have time to radiatively cool in the interval $t_{\rm wait}$ between flares.  

Consider first cooling due to synchrotron radiation.  The magnetic field strength of the wind, just upstream of the termination shock, is approximately
\begin{equation}
\left. B \right\vert_{R_{\rm n}} \approx 13\,{\rm G}\,B_{\rm d,15}^{-1}\left(\frac{t_{\rm age}}{30{\rm yr}}\right)^{-1}\,\left(\frac{R_{\rm n}}{10^{12}{\rm cm}}\right)^{-1},
\end{equation}
where we have used $L_{\rm sd} = (B|_{\rm R_{\rm n}}^{2}/8\pi)c\times 4\pi R_{\rm n}^{2}$ and eq.~(\ref{eq:Lsd}).  Pairs immediately behind the wind termination shock will cool, in a timescale $t_{\rm wait}$, to a Lorentz factor
\begin{eqnarray}
\gamma_{\rm c,syn} 
&\approx \frac{6\pi m_{\rm e} c}{\sigma_{\rm T}B|_{R_{\rm n}}^{2}t_{\rm wait}} \approx 4.6 \times 10^4 B_{\rm d,15}^{2}\left(\frac{t_{\rm age}}{30\,{\rm yr}}\right)^{2}\left(\frac{R_{\rm n}}{10^{12}\,{\rm cm}}\right)^{2}\left(\frac{t_{\rm wait}}{100{\rm s}}\right)^{-1} 
\nonumber \\
&\approx 3.4\times 10^{5} \left(\frac{t_{\rm age}}{30\,{\rm yr}}\right)^{6/5}B_{\rm d,15}^{6/5}\left(\frac{t_{\rm wait}}{100\,{\rm s}}\right)^{1/5}\left(\frac{n_{\rm sh}}{10^{4}\,{\rm cm^{-3}}}\right)^{-2/5}
\label{eq:gammasyn}
\end{eqnarray}
Typical values of $B_{\rm d}$ and $t_{\rm age}$ give $\gamma_{\rm c,syn} \gg 1$ , in which case the pairs will still be relativistically hot by the time of the next shock.

Another, potentially more important, cooling source is Compton scattering by photons generated by the shock's incoherent synchrotron radiation.  As discussed earlier, each FRB-producing shock will generate a pulse of photons of total energy $\sim E_{\rm flare}$ and duration $\delta t$ with a peak energy $h\nu_{\rm syn}$ (eq.~\ref{eq:nusyn}) in the optical or X-ray range if the upstream medium is highly loaded in $e^{-}/e^{+}$ pairs ($f_{\rm e} \gg 1$).  Provided that $h\nu_{\rm syn} \ll m_{\rm e} c^{2}/\gamma_{\pm} \sim 1\,{\rm keV}\,(\gamma_{\pm}/10^{3})^{-1}$, where $\gamma_{\pm}$ is the Lorentz factor to which synchrotron radiation has already cooled the pairs (eq.~\ref{eq:gammasyn}), then Klein-Nishina corrections to the electron scattering cross section are negligible.  Inverse Compton scattering by this photon pulse --- passing through the upstream electrons over a timescale $\delta t$ --- will then cool them to Lorentz factor
\begin{align} \label{eq:gammac}
\gamma_{\rm c, IC} 
&\approx \frac{3 m_{\rm e} c}{4\sigma_{\rm T}U_{\rm ph}\delta t} \approx \frac{6\pi m_{\rm e} c^2}{E_{\rm flare}}\frac{R_{\rm n}^{2}}{\sigma_{\rm T}}
\\ \nonumber
&\approx 1.7 E_{\rm flare,44}^{-1}\left(\frac{t_{\rm age}}{30\,{\rm yr}}\right)^{-4/5}B_{\rm d,15}^{-4/5}\left(\frac{t_{\rm wait}}{100{\,\rm s}}\right)^{6/5}\left(\frac{n_{\rm sh}}{10^{4}\,{\rm cm^{-3}}}\right)^{-2/5},
\end{align}
where $U_{\rm ph} = L_{\rm syn}/(4\pi R_{\rm n}^{2}c)$ is the energy density of the synchrotron photons at the termination shock, $L_{\rm syn} \simeq (0.5E_{\rm flare}/\delta t)$ is the peak synchrotron luminosity (acting over time $\delta t$), and in the final line we have used equation~(\ref{eq:Rneb}) for $R_{\rm n}$.  

Equation~(\ref{eq:gammac}) shows, that for magnetars which are sufficiently old or strongly magnetized $B_{\rm d} \gtrsim 10^{15}$ G, a particularly energetic flare $E_{\rm flare} \gtrsim 2 \times 10^{42}$ erg could act to cool the nebula pairs to mildly relativistic thermal velocities $\gamma \lesssim 1$, generating the cold upstream medium required for the operation of the synchrotron maser.  If such energetic flares were required to occur after {\it every} burst to keep the immediate upstream cold, this would strain the energy budget of the magnetar engine.    On the other hand, if such powerful cooling flares a rare, a large enough reservoir of ``cooled'' electrons could be maintained to allow for efficient maser emission.

\bibliographystyle{mnras}
\bibliography{refs.bib} 

\bsp	
\label{lastpage}
\end{document}